\documentclass[longauth]{aa}

\usepackage[varg]{txfonts}
\usepackage{eht}
\usepackage{multirow}
\usepackage{bm}
\usepackage{comment}
\usepackage{lastpage}

\usepackage[table]{xcolor}
\definecolor{tabgreen}{rgb}{0.88, 1, 0.88}
\newcommand{\ccg}{\cellcolor{tabgreen}}

\usepackage[symbol]{footmisc}
\renewcommand{\thefootnote}{\fnsymbol{footnote}}

\usepackage[breaklinks=true]{hyperref}
\hypersetup{
    colorlinks,
    linkcolor={blue!50!black},
    citecolor={blue!50!black},
    urlcolor={blue!50!black}
    }

\usepackage{orcidlink}
\let\orcid\orcidlink

\usepackage[encapsulated]{CJK}
\usepackage[utf8]{inputenc}
\DeclareUnicodeCharacter{02BC}{'}
\newcommand{\cntext}[1]{\begin{CJK}{UTF8}{gbsn}#1\end{CJK}}

\renewcommand{\linenumbers}{\relax}

\begin{document}
\makeatletter

\title{\vspace{-0.25cm}Origin of the ring ellipticity in the black hole images of M87*}

\author{\tiny
Rohan Dahale\orcid{0000-0001-6982-9034}\inst{\ref{1}}\thanks{These authors have contributed equally to this work.}\and
Ilje Cho\orcid{0000-0001-6083-7521}\inst{\ref{2},\ref{3},\ref{1} *}\and
Kotaro Moriyama \orcid{0000-0003-1364-3761} \inst{\ref{1}, \ref{4}} \and
Kaj Wiik \orcid{0000-0002-0862-3398} \inst{\ref{5}, \ref{6}, \ref{7}} \and
Paul Tiede \orcid{0000-0003-3826-5648} \inst{\ref{8}, \ref{9}} \and
José L. Gómez \orcid{0000-0003-4190-7613} \inst{\ref{1}} \and
Chi-kwan Chan \orcid{0000-0001-6337-6126} \inst{\ref{10}, \ref{11}, \ref{12}} \and
Roman Gold \orcid{0000-0003-2492-1966} \inst{\ref{13}, \ref{14}, \ref{15}} \and
Vadim Y. Bernshteyn \orcid{0009-0000-1376-2352} \inst{\ref{10}, \ref{12}} \and
Marianna Foschi \orcid{0000-0001-8147-4993} \inst{\ref{1}} \and
Britton Jeter \orcid{0000-0003-2847-1712} \inst{\ref{16}} \and
Hung-Yi Pu \orcid{0000-0001-9270-8812} \inst{\ref{17}, \ref{18}, \ref{16}} \and
Boris Georgiev \orcid{0000-0002-3586-6424} \inst{\ref{10}} \and
Abhishek V. Joshi \orcid{0000-0002-2514-5965} \inst{\ref{19}} \and
Alejandro Cruz-Osorio \orcid{0000-0002-3945-6342} \inst{\ref{20}, \ref{21}} \and
Iniyan Natarajan \orcid{0000-0001-8242-4373} \inst{\ref{8}, \ref{9}} \and
Avery E. Broderick \orcid{0000-0002-3351-760X} \inst{\ref{22}, \ref{23}, \ref{24}} \and
León D. S. Salas \orcid{0000-0003-1979-6363} \inst{\ref{25}} \and
Koushik Chatterjee \orcid{0000-0002-2825-3590} \inst{\ref{26}} \and
\\ ------------ \\
Kazunori Akiyama \orcid{0000-0002-9475-4254} \inst{\ref{27}, \ref{28}, \ref{9}} \and
Ezequiel Albentosa-Ruíz \orcid{0000-0002-7816-6401} \inst{\ref{29}} \and
Antxon Alberdi \orcid{0000-0002-9371-1033} \inst{\ref{1}} \and
Walter Alef \inst{\ref{30}} \and
Juan Carlos Algaba \orcid{0000-0001-6993-1696} \inst{\ref{31}} \and
Richard Anantua \orcid{0000-0003-3457-7660} \inst{\ref{32}, \ref{33}, \ref{9}, \ref{8}} \and
Keiichi Asada \orcid{0000-0001-6988-8763} \inst{\ref{16}} \and
Rebecca Azulay \orcid{0000-0002-2200-5393} \inst{\ref{29}, \ref{34}, \ref{30}} \and
Uwe Bach \orcid{0000-0002-7722-8412} \inst{\ref{30}} \and
Anne-Kathrin Baczko \orcid{0000-0003-3090-3975} \inst{\ref{35}, \ref{30}} \and
David Ball \inst{\ref{10}} \and
Mislav Baloković \orcid{0000-0003-0476-6647} \inst{\ref{36}} \and
Bidisha Bandyopadhyay \orcid{0000-0002-2138-8564} \inst{\ref{37}} \and
John Barrett \orcid{0000-0002-9290-0764} \inst{\ref{27}} \and
Michi Bauböck \orcid{0000-0002-5518-2812} \inst{\ref{19}} \and
Bradford A. Benson \orcid{0000-0002-5108-6823} \inst{\ref{38}, \ref{39}} \and
Dan Bintley \inst{\ref{40}, \ref{41}} \and
Lindy Blackburn \orcid{0000-0002-9030-642X} \inst{\ref{9}, \ref{8}} \and
Raymond Blundell \orcid{0000-0002-5929-5857} \inst{\ref{8}} \and
Katherine L. Bouman \orcid{0000-0003-0077-4367} \inst{\ref{42}} \and
Geoffrey C. Bower \orcid{0000-0003-4056-9982} \inst{\ref{43}, \ref{44}} \and
Michael Bremer \inst{\ref{45}} \and
Roger Brissenden \orcid{0000-0002-2556-0894} \inst{\ref{9}, \ref{8}} \and
Silke Britzen \orcid{0000-0001-9240-6734} \inst{\ref{30}} \and
Dominique Broguiere \orcid{0000-0001-9151-6683} \inst{\ref{45}} \and
Thomas Bronzwaer \orcid{0000-0003-1151-3971} \inst{\ref{46}} \and
Sandra Bustamante \orcid{0000-0001-6169-1894} \inst{\ref{47}} \and
Douglas Ferreira Carlos \orcid{0000-0002-1340-7702} \inst{\ref{48}} \and
John E. Carlstrom \orcid{0000-0002-2044-7665} \inst{\ref{49}, \ref{39}, \ref{50}, \ref{51}} \and
Andrew Chael \orcid{0000-0003-2966-6220} \inst{\ref{52}} \and
Dominic O. Chang \orcid{0000-0001-9939-5257} \inst{\ref{9}, \ref{8}} \and
Shami Chatterjee \orcid{0000-0002-2878-1502} \inst{\ref{53}} \and
Ming-Tang Chen \orcid{0000-0001-6573-3318} \inst{\ref{54}} \and
Yongjun Chen (\cntext{陈永军}) \orcid{0000-0001-5650-6770} \inst{\ref{55}, \ref{56}} \and
Xiaopeng Cheng \orcid{0000-0003-4407-9868} \inst{\ref{2}} \and
Pierre Christian \orcid{0000-0001-6820-9941} \inst{\ref{57}} \and
Nicholas S. Conroy \orcid{0000-0003-2886-2377} \inst{\ref{58}, \ref{8}} \and
John E. Conway \orcid{0000-0003-2448-9181} \inst{\ref{35}} \and
Thomas M. Crawford \orcid{0000-0001-9000-5013} \inst{\ref{39}, \ref{49}} \and
Geoffrey B. Crew \orcid{0000-0002-2079-3189} \inst{\ref{27}} \and
Yuzhu Cui (\cntext{崔玉竹}) \orcid{0000-0001-6311-4345} \inst{\ref{59}, \ref{60}} \and
Brandon Curd \orcid{0000-0002-8650-0879} \inst{\ref{32}, \ref{9}, \ref{8}} \and
Jordy Davelaar \orcid{0000-0002-2685-2434} \inst{\ref{61}, \ref{62}} \and
Mariafelicia De Laurentis \orcid{0000-0002-9945-682X} \inst{\ref{63}, \ref{64}} \and
Roger Deane \orcid{0000-0003-1027-5043} \inst{\ref{65}, \ref{66}, \ref{67}} \and
Jessica Dempsey \orcid{0000-0003-1269-9667} \inst{\ref{40}, \ref{41}, \ref{68}} \and
Gregory Desvignes \orcid{0000-0003-3922-4055} \inst{\ref{30}, \ref{69}} \and
Jason Dexter \orcid{0000-0003-3903-0373} \inst{\ref{70}} \and
Vedant Dhruv \orcid{0000-0001-6765-877X} \inst{\ref{19}} \and
Indu K. Dihingia \orcid{0000-0002-4064-0446} \inst{\ref{60}} \and
Sheperd S. Doeleman \orcid{0000-0002-9031-0904} \inst{\ref{9}, \ref{8}} \and
Sergio A. Dzib \orcid{0000-0001-6010-6200} \inst{\ref{30}} \and
Ralph P. Eatough \orcid{0000-0001-6196-4135} \inst{\ref{71}, \ref{30}} \and
Razieh Emami \orcid{0000-0002-2791-5011} \inst{\ref{8}} \and
Heino Falcke \orcid{0000-0002-2526-6724} \inst{\ref{46}} \and
Joseph Farah \orcid{0000-0003-4914-5625} \inst{\ref{72}, \ref{73}} \and
Vincent L. Fish \orcid{0000-0002-7128-9345} \inst{\ref{27}} \and
Edward Fomalont \orcid{0000-0002-9036-2747} \inst{\ref{74}} \and
H. Alyson Ford \orcid{0000-0002-9797-0972} \inst{\ref{10}} \and
Raquel Fraga-Encinas \orcid{0000-0002-5222-1361} \inst{\ref{46}} \and
William T. Freeman \inst{\ref{75}, \ref{76}} \and
Per Friberg \orcid{0000-0002-8010-8454} \inst{\ref{40}, \ref{41}} \and
Christian M. Fromm \orcid{0000-0002-1827-1656} \inst{\ref{77}, \ref{21}, \ref{30}} \and
Antonio Fuentes \orcid{0000-0002-8773-4933} \inst{\ref{1}} \and
Peter Galison \orcid{0000-0002-6429-3872} \inst{\ref{9}, \ref{78}, \ref{79}} \and
Charles F. Gammie \orcid{0000-0001-7451-8935} \inst{\ref{19}, \ref{58}, \ref{80}} \and
Roberto García \orcid{0000-0002-6584-7443} \inst{\ref{45}} \and
Olivier Gentaz \orcid{0000-0002-0115-4605} \inst{\ref{45}} \and
Gertie Geertsema \orcid{0000-0003-3933-0069} \inst{\ref{81}} \and
Ciriaco Goddi \orcid{0000-0002-2542-7743} \inst{\ref{82}, \ref{83}, \ref{84}, \ref{85}} \and
Arturo I. Gómez-Ruiz \orcid{0000-0001-9395-1670} \inst{\ref{86}, \ref{87}} \and
Minfeng Gu (\cntext{顾敏峰}) \orcid{0000-0002-4455-6946} \inst{\ref{55}, \ref{88}} \and
Mark Gurwell \orcid{0000-0003-0685-3621} \inst{\ref{8}} \and
Kazuhiro Hada \orcid{0000-0001-6906-772X} \inst{\ref{89}, \ref{4}} \and
Daryl Haggard \orcid{0000-0001-6803-2138} \inst{\ref{90}, \ref{91}} \and
Ronald Hesper \orcid{0000-0003-1918-6098} \inst{\ref{92}} \and
Dirk Heumann \orcid{0000-0002-7671-0047} \inst{\ref{10}} \and
Luis C. Ho (\cntext{何子山}) \orcid{0000-0001-6947-5846} \inst{\ref{93}, \ref{94}} \and
Paul Ho \orcid{0000-0002-3412-4306} \inst{\ref{16}, \ref{41}, \ref{40}} \and
Mareki Honma \orcid{0000-0003-4058-9000} \inst{\ref{4}, \ref{95}, \ref{96}} \and
Chih-Wei L. Huang \orcid{0000-0001-5641-3953} \inst{\ref{16}} \and
Lei Huang (\cntext{黄磊}) \orcid{0000-0002-1923-227X} \inst{\ref{55}, \ref{88}} \and
David H. Hughes \inst{\ref{86}} \and
Shiro Ikeda \orcid{0000-0002-2462-1448} \inst{\ref{28}, \ref{97}, \ref{98}, \ref{99}} \and
C. M. Violette Impellizzeri \orcid{0000-0002-3443-2472} \inst{\ref{100}, \ref{74}} \and
Makoto Inoue \orcid{0000-0001-5037-3989} \inst{\ref{16}} \and
Sara Issaoun \orcid{0000-0002-5297-921X} \inst{\ref{8}, \ref{62}} \and
David J. James \orcid{0000-0001-5160-4486} \inst{\ref{101}, \ref{102}} \and
Buell T. Jannuzi \orcid{0000-0002-1578-6582} \inst{\ref{10}} \and
Michael Janssen \orcid{0000-0001-8685-6544} \inst{\ref{46}, \ref{30}} \and
Wu Jiang (\cntext{江悟}) \orcid{0000-0001-7369-3539} \inst{\ref{55}} \and
Alejandra Jiménez-Rosales \orcid{0000-0002-2662-3754} \inst{\ref{46}} \and
Michael D. Johnson \orcid{0000-0002-4120-3029} \inst{\ref{9}, \ref{8}} \and
Svetlana Jorstad \orcid{0000-0001-6158-1708} \inst{\ref{103}} \and
Adam C. Jones \inst{\ref{39}} \and
Taehyun Jung \orcid{0000-0001-7003-8643} \inst{\ref{2}, \ref{104}} \and
Ramesh Karuppusamy \orcid{0000-0002-5307-2919} \inst{\ref{30}} \and
Tomohisa Kawashima \orcid{0000-0001-8527-0496} \inst{\ref{105}} \and
Garrett K. Keating \orcid{0000-0002-3490-146X} \inst{\ref{8}} \and
Mark Kettenis \orcid{0000-0002-6156-5617} \inst{\ref{106}} \and
Dong-Jin Kim \orcid{0000-0002-7038-2118} \inst{\ref{107}} \and
Jae-Young Kim \orcid{0000-0001-8229-7183} \inst{\ref{108}, \ref{30}} \and
Jongsoo Kim \orcid{0000-0002-1229-0426} \inst{\ref{2}} \and
Junhan Kim \orcid{0000-0002-4274-9373} \inst{\ref{109}} \and
Motoki Kino \orcid{0000-0002-2709-7338} \inst{\ref{28}, \ref{110}} \and
Jun Yi Koay \orcid{0000-0002-7029-6658} \inst{\ref{16}} \and
Prashant Kocherlakota \orcid{0000-0001-7386-7439} \inst{\ref{9}} \and
Yutaro Kofuji \inst{\ref{4}, \ref{96}} \and
Patrick M. Koch \orcid{0000-0003-2777-5861} \inst{\ref{16}} \and
Shoko Koyama \orcid{0000-0002-3723-3372} \inst{\ref{111}, \ref{16}} \and
Carsten Kramer \orcid{0000-0002-4908-4925} \inst{\ref{45}} \and
Joana A. Kramer \orcid{0009-0003-3011-0454} \inst{\ref{30}} \and
Michael Kramer \orcid{0000-0002-4175-2271} \inst{\ref{30}} \and
Thomas P. Krichbaum \orcid{0000-0002-4892-9586} \inst{\ref{30}} \and
Cheng-Yu Kuo \orcid{0000-0001-6211-5581} \inst{\ref{112}, \ref{16}} \and
Noemi La Bella \orcid{0000-0002-8116-9427} \inst{\ref{46}} \and
Sang-Sung Lee \orcid{0000-0002-6269-594X} \inst{\ref{2}} \and
Aviad Levis \orcid{0000-0001-7307-632X} \inst{\ref{42}} \and
Zhiyuan Li (\cntext{李志远}) \orcid{0000-0003-0355-6437} \inst{\ref{113}, \ref{114}} \and
Rocco Lico \orcid{0000-0001-7361-2460} \inst{\ref{115}, \ref{1}} \and
Greg Lindahl \orcid{0000-0002-6100-4772} \inst{\ref{116}} \and
Michael Lindqvist \orcid{0000-0002-3669-0715} \inst{\ref{35}} \and
Mikhail Lisakov \orcid{0000-0001-6088-3819} \inst{\ref{117}} \and
Jun Liu (\cntext{刘俊}) \orcid{0000-0002-7615-7499} \inst{\ref{30}} \and
Kuo Liu \orcid{0000-0002-2953-7376} \inst{\ref{55}, \ref{56}} \and
Elisabetta Liuzzo \orcid{0000-0003-0995-5201} \inst{\ref{118}} \and
Wen-Ping Lo \orcid{0000-0003-1869-2503} \inst{\ref{16}, \ref{119}} \and
Andrei P. Lobanov \orcid{0000-0003-1622-1484} \inst{\ref{30}} \and
Laurent Loinard \orcid{0000-0002-5635-3345} \inst{\ref{120}, \ref{9}, \ref{121}} \and
Colin J. Lonsdale \orcid{0000-0003-4062-4654} \inst{\ref{27}} \and
Amy E. Lowitz \orcid{0000-0002-4747-4276} \inst{\ref{10}} \and
Ru-Sen Lu (\cntext{路如森}) \orcid{0000-0002-7692-7967} \inst{\ref{55}, \ref{56}, \ref{30}} \and
Nicholas R. MacDonald \orcid{0000-0002-6684-8691} \inst{\ref{30}} \and
Jirong Mao (\cntext{毛基荣}) \orcid{0000-0002-7077-7195} \inst{\ref{122}, \ref{123}, \ref{124}} \and
Nicola Marchili \orcid{0000-0002-5523-7588} \inst{\ref{118}, \ref{30}} \and
Sera Markoff \orcid{0000-0001-9564-0876} \inst{\ref{25}, \ref{125}} \and
Daniel P. Marrone \orcid{0000-0002-2367-1080} \inst{\ref{10}} \and
Alan P. Marscher \orcid{0000-0001-7396-3332} \inst{\ref{103}} \and
Iván Martí-Vidal \orcid{0000-0003-3708-9611} \inst{\ref{29}, \ref{34}} \and
Satoki Matsushita \orcid{0000-0002-2127-7880} \inst{\ref{16}} \and
Lynn D. Matthews \orcid{0000-0002-3728-8082} \inst{\ref{27}} \and
Lia Medeiros \orcid{0000-0003-2342-6728} \inst{\ref{61}, \ref{62}} \and
Karl M. Menten \orcid{0000-0001-6459-0669} \inst{\ref{30}}\thanks{Deceased} \and
Izumi Mizuno \orcid{0000-0002-7210-6264} \inst{\ref{40}, \ref{41}} \and
Yosuke Mizuno \orcid{0000-0002-8131-6730} \inst{\ref{60}, \ref{127}, \ref{21}} \and
Joshua Montgomery \orcid{0000-0003-0345-8386} \inst{\ref{91}, \ref{39}} \and
James M. Moran \orcid{0000-0002-3882-4414} \inst{\ref{9}, \ref{8}} \and
Monika Moscibrodzka \orcid{0000-0002-4661-6332} \inst{\ref{46}} \and
Wanga Mulaudzi \orcid{0000-0003-4514-625X} \inst{\ref{25}} \and
Cornelia Müller \orcid{0000-0002-2739-2994} \inst{\ref{30}, \ref{46}} \and
Hendrik Müller \orcid{0000-0002-9250-0197} \inst{\ref{30}} \and
Alejandro Mus \orcid{0000-0003-0329-6874} \inst{\ref{83}, \ref{115}} \and
Gibwa Musoke \orcid{0000-0003-1984-189X} \inst{\ref{25}, \ref{46}} \and
Ioannis Myserlis \orcid{0000-0003-3025-9497} \inst{\ref{128}} \and
Hiroshi Nagai \orcid{0000-0003-0292-3645} \inst{\ref{28}, \ref{95}} \and
Neil M. Nagar \orcid{0000-0001-6920-662X} \inst{\ref{37}} \and
Dhanya G. Nair \orcid{0000-0001-5357-7805} \inst{\ref{37}, \ref{30}} \and
Masanori Nakamura \orcid{0000-0001-6081-2420} \inst{\ref{129}, \ref{16}} \and
Gopal Narayanan \orcid{0000-0002-4723-6569} \inst{\ref{47}} \and
Antonios Nathanail \orcid{0000-0002-1655-9912} \inst{\ref{130}, \ref{21}} \and
Santiago Navarro Fuentes \inst{\ref{128}} \and
Joey Neilsen \orcid{0000-0002-8247-786X} \inst{\ref{131}} \and
Chunchong Ni \orcid{0000-0003-1361-5699} \inst{\ref{23}, \ref{24}, \ref{22}} \and
Michael A. Nowak \orcid{0000-0001-6923-1315} \inst{\ref{132}} \and
Junghwan Oh \orcid{0000-0002-4991-9638} \inst{\ref{106}} \and
Hiroki Okino \orcid{0000-0003-3779-2016} \inst{\ref{4}, \ref{96}} \and
Héctor Raúl Olivares Sánchez \orcid{0000-0001-6833-7580} \inst{\ref{133}} \and
Tomoaki Oyama \orcid{0000-0003-4046-2923} \inst{\ref{4}} \and
Feryal Özel \orcid{0000-0003-4413-1523} \inst{\ref{134}} \and
Daniel C. M. Palumbo \orcid{0000-0002-7179-3816} \inst{\ref{9}, \ref{8}} \and
Georgios Filippos Paraschos \orcid{0000-0001-6757-3098} \inst{\ref{30}} \and
Jongho Park \orcid{0000-0001-6558-9053} \inst{\ref{135}, \ref{16}} \and
Harriet Parsons \orcid{0000-0002-6327-3423} \inst{\ref{40}, \ref{41}} \and
Nimesh Patel \orcid{0000-0002-6021-9421} \inst{\ref{8}} \and
Ue-Li Pen \orcid{0000-0003-2155-9578} \inst{\ref{16}, \ref{22}, \ref{136}, \ref{137}, \ref{138}} \and
Dominic W. Pesce \orcid{0000-0002-5278-9221} \inst{\ref{8}, \ref{9}} \and
Vincent Piétu \inst{\ref{45}} \and
Aleksandar PopStefanija \inst{\ref{47}} \and
Oliver Porth \orcid{0000-0002-4584-2557} \inst{\ref{25}, \ref{21}} \and
Ben Prather \orcid{0000-0002-0393-7734} \inst{\ref{19}} \and
Giacomo Principe \orcid{0000-0003-0406-7387} \inst{\ref{139}, \ref{140}, \ref{115}} \and
Dimitrios Psaltis \orcid{0000-0003-1035-3240} \inst{\ref{134}} \and
Venkatessh Ramakrishnan \orcid{0000-0002-9248-086X} \inst{\ref{37}, \ref{6}, \ref{7}} \and
Ramprasad Rao \orcid{0000-0002-1407-7944} \inst{\ref{8}} \and
Mark G. Rawlings \orcid{0000-0002-6529-202X} \inst{\ref{141}, \ref{40}, \ref{41}} \and
Luciano Rezzolla \orcid{0000-0002-1330-7103} \inst{\ref{21}, \ref{142}, \ref{143}} \and
Angelo Ricarte \orcid{0000-0001-5287-0452} \inst{\ref{9}, \ref{8}} \and
Bart Ripperda \orcid{0000-0002-7301-3908} \inst{\ref{136}, \ref{144}, \ref{137}, \ref{22}} \and
Jan Röder \orcid{0000-0002-2426-927X} \inst{\ref{1}} \and
Freek Roelofs \orcid{0000-0001-5461-3687} \inst{\ref{46}} \and
Cristina Romero-Cañizales \orcid{0000-0001-6301-9073} \inst{\ref{16}} \and
Eduardo Ros \orcid{0000-0001-9503-4892} \inst{\ref{30}} \and
Arash Roshanineshat \orcid{0000-0002-8280-9238} \inst{\ref{10}} \and
Helge Rottmann \inst{\ref{30}} \and
Alan L. Roy \orcid{0000-0002-1931-0135} \inst{\ref{30}} \and
Ignacio Ruiz \orcid{0000-0002-0965-5463} \inst{\ref{128}} \and
Chet Ruszczyk \orcid{0000-0001-7278-9707} \inst{\ref{27}} \and
Kazi L. J. Rygl \orcid{0000-0003-4146-9043} \inst{\ref{118}} \and
Salvador Sánchez \orcid{0000-0002-8042-5951} \inst{\ref{128}} \and
David Sánchez-Argüelles \orcid{0000-0002-7344-9920} \inst{\ref{86}, \ref{87}} \and
Miguel Sánchez-Portal \orcid{0000-0003-0981-9664} \inst{\ref{128}} \and
Mahito Sasada \orcid{0000-0001-5946-9960} \inst{\ref{145}, \ref{4}, \ref{146}} \and
Kaushik Satapathy \orcid{0000-0003-0433-3585} \inst{\ref{10}} \and
Saurabh \orcid{0000-0001-7156-4848} \inst{\ref{30}} \and
Tuomas Savolainen \orcid{0000-0001-6214-1085} \inst{\ref{147}, \ref{7}, \ref{30}} \and
F. Peter Schloerb \inst{\ref{47}} \and
Jonathan Schonfeld \orcid{0000-0002-8909-2401} \inst{\ref{8}} \and
Karl-Friedrich Schuster \orcid{0000-0003-2890-9454} \inst{\ref{148}} \and
Lijing Shao \orcid{0000-0002-1334-8853} \inst{\ref{94}, \ref{30}} \and
Zhiqiang Shen (\cntext{沈志强}) \orcid{0000-0003-3540-8746} \inst{\ref{55}, \ref{56}} \and
Sasikumar Silpa \orcid{0000-0003-0667-7074} \inst{\ref{37}} \and
Des Small \orcid{0000-0003-3723-5404} \inst{\ref{106}} \and
Bong Won Sohn \orcid{0000-0002-4148-8378} \inst{\ref{2}, \ref{104}, \ref{3}} \and
Jason SooHoo \orcid{0000-0003-1938-0720} \inst{\ref{27}} \and
Kamal Souccar \orcid{0000-0001-7915-5272} \inst{\ref{47}} \and
Joshua S. Stanway \orcid{0009-0003-7659-4642} \inst{\ref{149}} \and
He Sun (\cntext{孙赫}) \orcid{0000-0003-1526-6787} \inst{\ref{150}, \ref{151}} \and
Fumie Tazaki \orcid{0000-0003-0236-0600} \inst{\ref{152}} \and
Alexandra J. Tetarenko \orcid{0000-0003-3906-4354} \inst{\ref{153}} \and
Remo P. J. Tilanus \orcid{0000-0002-6514-553X} \inst{\ref{10}, \ref{46}, \ref{100}, \ref{154}} \and
Michael Titus \orcid{0000-0001-9001-3275} \inst{\ref{27}} \and
Kenji Toma \orcid{0000-0002-7114-6010} \inst{\ref{155}, \ref{156}} \and
Pablo Torne \orcid{0000-0001-8700-6058} \inst{\ref{128}, \ref{30}} \and
Teresa Toscano \orcid{0000-0003-3658-7862} \inst{\ref{1}} \and
Efthalia Traianou \orcid{0000-0002-1209-6500} \inst{\ref{1}, \ref{30}} \and
Tyler Trent \inst{\ref{10}} \and
Sascha Trippe \orcid{0000-0003-0465-1559} \inst{\ref{157}} \and
Matthew Turk \orcid{0000-0002-5294-0198} \inst{\ref{58}} \and
Ilse van Bemmel \orcid{0000-0001-5473-2950} \inst{\ref{68}} \and
Huib Jan van Langevelde \orcid{0000-0002-0230-5946} \inst{\ref{106}, \ref{100}, \ref{158}} \and
Daniel R. van Rossum \orcid{0000-0001-7772-6131} \inst{\ref{46}} \and
Jesse Vos \orcid{0000-0003-3349-7394} \inst{\ref{46}} \and
Jan Wagner \orcid{0000-0003-1105-6109} \inst{\ref{30}} \and
Derek Ward-Thompson \orcid{0000-0003-1140-2761} \inst{\ref{149}} \and
John Wardle \orcid{0000-0002-8960-2942} \inst{\ref{159}} \and
Jasmin E. Washington \orcid{0000-0002-7046-0470} \inst{\ref{10}} \and
Jonathan Weintroub \orcid{0000-0002-4603-5204} \inst{\ref{9}, \ref{8}} \and
Robert Wharton \orcid{0000-0002-7416-5209} \inst{\ref{30}} \and
Maciek Wielgus \orcid{0000-0002-8635-4242} \inst{\ref{1}} \and
Gunther Witzel \orcid{0000-0003-2618-797X} \inst{\ref{30}} \and
Michael F. Wondrak \orcid{0000-0002-6894-1072} \inst{\ref{46}, \ref{160}} \and
George N. Wong \orcid{0000-0001-6952-2147} \inst{\ref{161}, \ref{52}} \and
Qingwen Wu (\cntext{吴庆文}) \orcid{0000-0003-4773-4987} \inst{\ref{162}} \and
Nitika Yadlapalli \orcid{0000-0003-3255-4617} \inst{\ref{42}} \and
Paul Yamaguchi \orcid{0000-0002-6017-8199} \inst{\ref{8}} \and
Aristomenis Yfantis \orcid{0000-0002-3244-7072} \inst{\ref{46}} \and
Doosoo Yoon \orcid{0000-0001-8694-8166} \inst{\ref{25}} \and
André Young \orcid{0000-0003-0000-2682} \inst{\ref{46}} \and
Ziri Younsi \orcid{0000-0001-9283-1191} \inst{\ref{163}, \ref{21}} \and
Wei Yu (\cntext{于威}) \orcid{0000-0002-5168-6052} \inst{\ref{8}} \and
Feng Yuan (\cntext{袁峰}) \orcid{0000-0003-3564-6437} \inst{\ref{164}} \and
Ye-Fei Yuan (\cntext{袁业飞}) \orcid{0000-0002-7330-4756} \inst{\ref{165}} \and
Ai-Ling Zeng (\cntext{曾艾玲}) \orcid{0009-0000-9427-4608} \inst{\ref{1}} \and
J. Anton Zensus \orcid{0000-0001-7470-3321} \inst{\ref{30}} \and
Shuo Zhang \orcid{0000-0002-2967-790X} \inst{\ref{166}} \and
Guang-Yao Zhao \orcid{0000-0002-4417-1659} \inst{\ref{1}, \ref{30}} \and
Shan-Shan Zhao (\cntext{赵杉杉}) \orcid{0000-0002-9774-3606} \inst{\ref{55}}
}

\institute{
Instituto de Astrofísica de Andalucía-CSIC, Glorieta de la Astronomía s/n, E-18008 Granada, Spain, \email{astrorohandahale@gmail.com} \label{1}  \and
Korea Astronomy and Space Science Institute, Daedeok-daero 776, Yuseong-gu, Daejeon 34055, Republic of Korea \label{2} \and
Department of Astronomy, Yonsei University, Yonsei-ro 50, Seodaemun-gu, 03722 Seoul, Republic of Korea \label{3} \and
Mizusawa VLBI Observatory, National Astronomical Observatory of Japan, 2-12 Hoshigaoka, Mizusawa, Oshu, Iwate 023-0861, Japan \label{4} \and
Tuorla Observatory, Department of Physics and Astronomy, University of Turku, FI-20014 Turun Yliopisto, Finland \label{5} \and
Finnish Centre for Astronomy with ESO, University of Turku, FI-20014 Turun Yliopisto, Finland \label{6} \and
Aalto University Metsähovi Radio Observatory, Metsähovintie 114, FI-02540 Kylmälä, Finland \label{7} \and
Center for Astrophysics $|$ Harvard \& Smithsonian, 60 Garden Street, Cambridge, MA 02138, USA \label{8} \and
Black Hole Initiative at Harvard University, 20 Garden Street, Cambridge, MA 02138, USA \label{9} \and
Steward Observatory and Department of Astronomy, University of Arizona, 933 N. Cherry Ave., Tucson, AZ 85721, USA \label{10} \and
Data Science Institute, University of Arizona, 1230 N. Cherry Ave., Tucson, AZ 85721, USA \label{11} \and
Program in Applied Mathematics, University of Arizona, 617 N. Santa Rita, Tucson, AZ 85721, USA \label{12} \and
Institute for Mathematics and Interdisciplinary Center for Scientific Computing, Heidelberg University, Im Neuenheimer Feld 205, Heidelberg 69120, Germany \label{13} \and
Institut f\"ur Theoretische Physik, Universit\"at Heidelberg, Philosophenweg 16, 69120 Heidelberg, Germany \label{14} \and
CP3-Origins, University of Southern Denmark, Campusvej 55, DK-5230 Odense, Denmark \label{15} \and
Institute of Astronomy and Astrophysics, Academia Sinica, 11F of Astronomy-Mathematics Building, AS/NTU No. 1, Sec. 4, Roosevelt Rd., Taipei 106216, Taiwan, R.O.C. \label{16} \and
Department of Physics, National Taiwan Normal University, No. 88, Sec. 4, Tingzhou Rd., Taipei 116, Taiwan, R.O.C. \label{17} \and
Center of Astronomy and Gravitation, National Taiwan Normal University, No. 88, Sec. 4, Tingzhou Road, Taipei 116, Taiwan, R.O.C. \label{18} \and
Department of Physics, University of Illinois, 1110 West Green Street, Urbana, IL 61801, USA \label{19} \and
Instituto de Astronomía, Universidad Nacional Autónoma de México (UNAM), Apdo Postal 70-264, Ciudad de México, México \label{20} \and
Institut für Theoretische Physik, Goethe-Universität Frankfurt, Max-von-Laue-Straße 1, D-60438 Frankfurt am Main, Germany \label{21} \and
Perimeter Institute for Theoretical Physics, 31 Caroline Street North, Waterloo, ON N2L 2Y5, Canada \label{22} \and
Department of Physics and Astronomy, University of Waterloo, 200 University Avenue West, Waterloo, ON N2L 3G1, Canada \label{23} \and
Waterloo Centre for Astrophysics, University of Waterloo, Waterloo, ON N2L 3G1, Canada \label{24} \and
Anton Pannekoek Institute for Astronomy, University of Amsterdam, Science Park 904, 1098 XH, Amsterdam, The Netherlands \label{25} \and
Institute for Research in Electronics and Applied Physics, University of Maryland, 8279 Paint Branch Drive, College Park, MD 20742, USA \label{26} \and
Massachusetts Institute of Technology Haystack Observatory, 99 Millstone Road, Westford, MA 01886, USA \label{27} \and
National Astronomical Observatory of Japan, 2-21-1 Osawa, Mitaka, Tokyo 181-8588, Japan \label{28} \and
Departament d'Astronomia i Astrofísica, Universitat de València, C. Dr. Moliner 50, E-46100 Burjassot, València, Spain \label{29} \and
Max-Planck-Institut für Radioastronomie, Auf dem Hügel 69, D-53121 Bonn, Germany \label{30} \and
Department of Physics, Faculty of Science, Universiti Malaya, 50603 Kuala Lumpur, Malaysia \label{31} \and
Department of Physics \& Astronomy, The University of Texas at San Antonio, One UTSA Circle, San Antonio, TX 78249, USA \label{32} \and
Physics \& Astronomy Department, Rice University, Houston, TX 77005-1827, USA \label{33} \and
Observatori Astronòmic, Universitat de València, C. Catedrático José Beltrán 2, E-46980 Paterna, València, Spain \label{34} \and
Department of Space, Earth and Environment, Chalmers University of Technology, Onsala Space Observatory, SE-43992 Onsala, Sweden \label{35} \and
Yale Center for Astronomy \& Astrophysics, Yale University, 52 Hillhouse Avenue, New Haven, CT 06511, USA \label{36} \and
Astronomy Department, Universidad de Concepción, Casilla 160-C, Concepción, Chile \label{37} \and
Fermi National Accelerator Laboratory, MS209, P.O. Box 500, Batavia, IL 60510, USA \label{38} \and
Department of Astronomy and Astrophysics, University of Chicago, 5640 South Ellis Avenue, Chicago, IL 60637, USA \label{39} \and
East Asian Observatory, 660 N. A'ohoku Place, Hilo, HI 96720, USA \label{40} \and
James Clerk Maxwell Telescope (JCMT), 660 N. A'ohoku Place, Hilo, HI 96720, USA \label{41} \and
California Institute of Technology, 1200 East California Boulevard, Pasadena, CA 91125, USA \label{42} \and
Institute of Astronomy and Astrophysics, Academia Sinica, \label{43} \and
Department of Physics and Astronomy, University of Hawaii at Manoa, 2505 Correa Road, Honolulu, HI 96822, USA \label{44} \and
Institut de Radioastronomie Millimétrique (IRAM), 300 rue de la Piscine, F-38406 Saint Martin d'Hères, France \label{45} \and
Department of Astrophysics, Institute for Mathematics, Astrophysics and Particle Physics (IMAPP), Radboud University, P.O. Box 9010, 6500 GL Nijmegen, The Netherlands \label{46} \and
Department of Astronomy, University of Massachusetts, Amherst, MA 01003, USA \label{47} \and
Instituto de Astronomia, Geofísica e Ci\^{e}ncias Atmosf\'{e}ricas da Universidade de S\~{a}o Paulo, Brazil \label{48} \and
Kavli Institute for Cosmological Physics, University of Chicago, 5640 South Ellis Avenue, Chicago, IL 60637, USA \label{49} \and
Department of Physics, University of Chicago, 5720 South Ellis Avenue, Chicago, IL 60637, USA \label{50} \and
Enrico Fermi Institute, University of Chicago, 5640 South Ellis Avenue, Chicago, IL 60637, USA \label{51} \and
Princeton Gravity Initiative, Jadwin Hall, Princeton University, Princeton, NJ 08544, USA \label{52} \and
Cornell Center for Astrophysics and Planetary Science, Cornell University, Ithaca, NY 14853, USA \label{53} \and
Institute of Astronomy and Astrophysics, Academia Sinica, 645 N. A'ohoku Place, Hilo, HI 96720, USA \label{54} \and
Shanghai Astronomical Observatory, Chinese Academy of Sciences, 80 Nandan Road, Shanghai 200030, People's Republic of China \label{55} \and
Key Laboratory of Radio Astronomy and Technology, Chinese Academy of Sciences, A20 Datun Road, Chaoyang District, Beijing, 100101, People’s Republic of China \label{56} \and
WattTime, 490 43rd Street, Unit 221, Oakland, CA 94609, USA \label{57} \and
Department of Astronomy, University of Illinois at Urbana-Champaign, 1002 West Green Street, Urbana, IL 61801, USA \label{58} \and
Research Center for Astronomical Computing, Zhejiang Laboratory, Hangzhou 311100, Peopleʼs Republic of China \label{59} \and
Tsung-Dao Lee Institute, Shanghai Jiao Tong University, Shengrong Road 520, Shanghai, 201210, People’s Republic of China \label{60} \and
Department of Astrophysical Sciences, Peyton Hall, Princeton University, Princeton, NJ 08544, USA \label{61} \and
NASA Hubble Fellowship Program, Einstein Fellow \label{62} \and
Dipartimento di Fisica ``E. Pancini'', Università di Napoli ``Federico II'', Compl. Univ. di Monte S. Angelo, Edificio G, Via Cinthia, I-80126, Napoli, Italy \label{63} \and
INFN Sez. di Napoli, Compl. Univ. di Monte S. Angelo, Edificio G, Via Cinthia, I-80126, Napoli, Italy \label{64} \and
Wits Centre for Astrophysics, University of the Witwatersrand, 1 Jan Smuts Avenue, Braamfontein, Johannesburg 2050, South Africa \label{65} \and
Department of Physics, University of Pretoria, Hatfield, Pretoria 0028, South Africa \label{66} \and
Centre for Radio Astronomy Techniques and Technologies, Department of Physics and Electronics, Rhodes University, Makhanda 6140, South Africa \label{67} \and
ASTRON, Oude Hoogeveensedijk 4, 7991 PD Dwingeloo, The Netherlands \label{68} \and
LESIA, Observatoire de Paris, Université PSL, CNRS, Sorbonne Université, Université de Paris, 5 place Jules Janssen, F-92195 Meudon, France \label{69} \and
JILA and Department of Astrophysical and Planetary Sciences, University of Colorado, Boulder, CO 80309, USA \label{70} \and
National Astronomical Observatories, Chinese Academy of Sciences, 20A Datun Road, Chaoyang District, Beijing 100101, PR China \label{71} \and
Las Cumbres Observatory, 6740 Cortona Drive, Suite 102, Goleta, CA 93117-5575, USA \label{72} \and
Department of Physics, University of California, Santa Barbara, CA 93106-9530, USA \label{73} \and
National Radio Astronomy Observatory, 520 Edgemont Road, Charlottesville, \label{74} \and
Department of Electrical Engineering and Computer Science, Massachusetts Institute of Technology, 32-D476, 77 Massachusetts Ave., Cambridge, MA 02142, USA \label{75} \and
Google Research, 355 Main St., Cambridge, MA 02142, USA \label{76} \and
Institut für Theoretische Physik und Astrophysik, Universität Würzburg, Emil-Fischer-Str. 31, \label{77} \and
Department of History of Science, Harvard University, Cambridge, MA 02138, USA \label{78} \and
Department of Physics, Harvard University, Cambridge, MA 02138, USA \label{79} \and
NCSA, University of Illinois, 1205 W. Clark St., Urbana, IL 61801, USA \label{80} \and
Royal Netherlands Meteorological Institute, Utrechtseweg 297, 3731 GA, De Bilt, The Netherlands \label{81} \and
Instituto de Astronomia, Geofísica e Ciências Atmosféricas, Universidade de São Paulo, R. do Matão, 1226, São Paulo, SP 05508-090, Brazil \label{82} \and
Dipartimento di Fisica, Università degli Studi di Cagliari, SP Monserrato-Sestu km 0.7, I-09042 Monserrato (CA), Italy \label{83} \and
INAF - Osservatorio Astronomico di Cagliari, via della Scienza 5, I-09047 Selargius (CA), Italy \label{84} \and
INFN, sezione di Cagliari, I-09042 Monserrato (CA), Italy \label{85} \and
Instituto Nacional de Astrofísica, Óptica y Electrónica. Apartado Postal 51 y 216, 72000. Puebla Pue., México \label{86} \and
Consejo Nacional de Humanidades, Ciencia y Tecnología, Av. Insurgentes Sur 1582, 03940, Ciudad de México, México \label{87} \and
Key Laboratory for Research in Galaxies and Cosmology, Chinese Academy of Sciences, Shanghai 200030, People's Republic of China \label{88} \and
Graduate School of Science, Nagoya City University, Yamanohata 1, Mizuho-cho, Mizuho-ku, Nagoya, 467-8501, Aichi, Japan \label{89} \and
Department of Physics, McGill University, 3600 rue University, Montréal, QC H3A 2T8, Canada \label{90} \and
Trottier Space Institute at McGill, 3550 rue University, Montréal,  QC H3A 2A7, Canada \label{91} \and
NOVA Sub-mm Instrumentation Group, Kapteyn Astronomical Institute, University of Groningen, Landleven 12, 9747 AD Groningen, The Netherlands \label{92} \and
Department of Astronomy, School of Physics, Peking University, Beijing 100871, People's Republic of China \label{93} \and
Kavli Institute for Astronomy and Astrophysics, Peking University, Beijing 100871, People's Republic of China \label{94} \and
Department of Astronomical Science, The Graduate University for Advanced Studies (SOKENDAI), 2-21-1 Osawa, Mitaka, Tokyo 181-8588, Japan \label{95} \and
Department of Astronomy, Graduate School of Science, The University of Tokyo, 7-3-1 Hongo, Bunkyo-ku, Tokyo 113-0033, Japan \label{96} \and
The Institute of Statistical Mathematics, 10-3 Midori-cho, Tachikawa, Tokyo, 190-8562, Japan \label{97} \and
Department of Statistical Science, The Graduate University for Advanced Studies (SOKENDAI), 10-3 Midori-cho, Tachikawa, Tokyo 190-8562, Japan \label{98} \and
Kavli Institute for the Physics and Mathematics of the Universe, The University of Tokyo, 5-1-5 Kashiwanoha, Kashiwa, 277-8583, Japan \label{99} \and
Leiden Observatory, Leiden University, Postbus 2300, 9513 RA Leiden, The Netherlands \label{100} \and
ASTRAVEO LLC, PO Box 1668, Gloucester, MA 01931, USA \label{101} \and
Applied Materials Inc., 35 Dory Road, Gloucester, MA 01930, USA \label{102} \and
Institute for Astrophysical Research, Boston University, 725 Commonwealth Ave., Boston, MA 02215, USA \label{103} \and
University of Science and Technology, Gajeong-ro 217, Yuseong-gu, Daejeon 34113, Republic of Korea \label{104} \and
Institute for Cosmic Ray Research, The University of Tokyo, 5-1-5 Kashiwanoha, Kashiwa, Chiba 277-8582, Japan \label{105} \and
Joint Institute for VLBI ERIC (JIVE), Oude Hoogeveensedijk 4, 7991 PD Dwingeloo, The Netherlands \label{106} \and
CSIRO, Space and Astronomy, PO Box 76, Epping, NSW 1710, Australia \label{107} \and
Department of Physics, Ulsan National Institute of Science and Technology (UNIST), Ulsan 44919, Republic of Korea \label{108} \and
Department of Physics, Korea Advanced Institute of Science and Technology (KAIST), 291 Daehak-ro, Yuseong-gu, Daejeon 34141, Republic of Korea \label{109} \and
Kogakuin University of Technology \& Engineering, Academic Support Center, 2665-1 Nakano, Hachioji, Tokyo 192-0015, Japan \label{110} \and
Graduate School of Science and Technology, Niigata University, 8050 Ikarashi 2-no-cho, Nishi-ku, Niigata 950-2181, Japan \label{111} \and
Physics Department, National Sun Yat-Sen University, No. 70, Lien-Hai Road, Kaosiung City 80424, Taiwan, R.O.C. \label{112} \and
School of Astronomy and Space Science, Nanjing University, Nanjing 210023, People's Republic of China \label{113} \and
Key Laboratory of Modern Astronomy and Astrophysics, Nanjing University, Nanjing 210023, People's Republic of China \label{114} \and
INAF-Istituto di Radioastronomia, Via P. Gobetti 101, I-40129 Bologna, Italy \label{115} \and
Common Crawl Foundation, 9663 Santa Monica Blvd. 425, Beverly Hills, CA 90210 USA \label{116} \and
Instituto de Física, Pontificia Universidad Católica de Valparaíso, Casilla 4059, Valparaíso, Chile \label{117} \and
INAF-Istituto di Radioastronomia \& Italian ALMA Regional Centre, Via P. Gobetti 101, I-40129 Bologna, Italy \label{118} \and
Department of Physics, National Taiwan University, No. 1, Sec. 4, Roosevelt Rd., Taipei 106216, Taiwan, R.O.C \label{119} \and
Instituto de Radioastronomía y Astrofísica, Universidad Nacional Autónoma de México, Morelia 58089, México \label{120} \and
David Rockefeller Center for Latin American Studies, Harvard University, 1730 Cambridge Street, Cambridge, MA 02138, USA \label{121} \and
Yunnan Observatories, Chinese Academy of Sciences, 650011 Kunming, Yunnan Province, People's Republic of China \label{122} \and
Center for Astronomical Mega-Science, Chinese Academy of Sciences, 20A Datun Road, Chaoyang District, Beijing, 100012, People's Republic of China \label{123} \and
Key Laboratory for the Structure and Evolution of Celestial Objects, Chinese Academy of Sciences, 650011 Kunming, People's Republic of China \label{124} \and
Gravitation and Astroparticle Physics Amsterdam (GRAPPA) Institute, University of Amsterdam, Science Park 904, 1098 XH Amsterdam, The Netherlands \label{125} \and
Deceased \label{126} \and
School of Physics and Astronomy, Shanghai Jiao Tong University, \label{127} \and
Institut de Radioastronomie Millimétrique (IRAM), Avenida Divina Pastora 7, Local 20, E-18012, Granada, Spain \label{128} \and
National Institute of Technology, Hachinohe College, 16-1 Uwanotai, Tamonoki, Hachinohe City, Aomori 039-1192, Japan \label{129} \and
Research Center for Astronomy, Academy of Athens, Soranou Efessiou 4, 115 27 Athens, Greece \label{130} \and
Department of Physics, Villanova University, 800 Lancaster Avenue, Villanova, PA 19085, USA \label{131} \and
Physics Department, Washington University, CB 1105, St. Louis, MO 63130, USA \label{132} \and
Departamento de Matemática da Universidade de Aveiro and Centre for Research and Development in Mathematics and Applications (CIDMA), Campus de Santiago, 3810-193 Aveiro, Portugal \label{133} \and
School of Physics, Georgia Institute of Technology, 837 State St NW, Atlanta, GA 30332, USA \label{134} \and
School of Space Research, Kyung Hee University, 1732, Deogyeong-daero, Giheung-gu, Yongin-si, Gyeonggi-do 17104, Republic of Korea \label{135} \and
Canadian Institute for Theoretical Astrophysics, University of Toronto, 60 St. George Street, Toronto, ON M5S 3H8, Canada \label{136} \and
Dunlap Institute for Astronomy and Astrophysics, University of Toronto, 50 St. George Street, Toronto, ON M5S 3H4, Canada \label{137} \and
Canadian Institute for Advanced Research, 180 Dundas St West, Toronto, ON M5G 1Z8, Canada \label{138} \and
Dipartimento di Fisica, Università di Trieste, I-34127 Trieste, Italy \label{139} \and
INFN Sez. di Trieste, I-34127 Trieste, Italy \label{140} \and
Gemini Observatory/NSF NOIRLab, 670 N. A’ohōkū Place, Hilo, HI 96720, USA \label{141} \and
Frankfurt Institute for Advanced Studies, Ruth-Moufang-Strasse 1, D-60438 Frankfurt, Germany \label{142} \and
School of Mathematics, Trinity College, Dublin 2, Ireland \label{143} \and
Department of Physics, University of Toronto, 60 St. George Street, Toronto, ON M5S 1A7, Canada \label{144} \and
Department of Physics, Tokyo Institute of Technology, 2-12-1 Ookayama, Meguro-ku, Tokyo 152-8551, Japan \label{145} \and
Hiroshima Astrophysical Science Center, Hiroshima University, 1-3-1 Kagamiyama, Higashi-Hiroshima, Hiroshima 739-8526, Japan \label{146} \and
Aalto University Department of Electronics and Nanoengineering, PL 15500, FI-00076 Aalto, Finland \label{147} \and
Institut de Radioastronomie Millimétrique (IRAM), 300 rue de la Piscine, \label{148} \and
Jeremiah Horrocks Institute, University of Central Lancashire, Preston PR1 2HE, UK \label{149} \and
National Biomedical Imaging Center, Peking University, Beijing 100871, People’s Republic of China \label{150} \and
College of Future Technology, Peking University, Beijing 100871, People’s Republic of China \label{151} \and
Tokyo Electron Technology Solutions Limited, 52 Matsunagane, Iwayado, Esashi, Oshu, Iwate 023-1101, Japan \label{152} \and
Department of Physics and Astronomy, University of Lethbridge, Lethbridge, Alberta T1K 3M4, Canada \label{153} \and
Netherlands Organisation for Scientific Research (NWO), Postbus 93138, 2509 AC Den Haag, The Netherlands \label{154} \and
Frontier Research Institute for Interdisciplinary Sciences, Tohoku University, Sendai 980-8578, Japan \label{155} \and
Astronomical Institute, Tohoku University, Sendai 980-8578, Japan \label{156} \and
Department of Physics and Astronomy, Seoul National University, Gwanak-gu, Seoul 08826, Republic of Korea \label{157} \and
University of New Mexico, Department of Physics and Astronomy, Albuquerque, NM 87131, USA \label{158} \and
Physics Department, Brandeis University, 415 South Street, Waltham, MA 02453, USA \label{159} \and
Radboud Excellence Fellow of Radboud University, Nijmegen, The Netherlands \label{160} \and
School of Natural Sciences, Institute for Advanced Study, 1 Einstein Drive, Princeton, NJ 08540, USA \label{161} \and
School of Physics, Huazhong University of Science and Technology, Wuhan, Hubei, 430074, People's Republic of China \label{162} \and
Mullard Space Science Laboratory, University College London, Holmbury St. Mary, Dorking, Surrey, RH5 6NT, UK \label{163} \and
Center for Astronomy and Astrophysics and Department of Physics, Fudan University, Shanghai 200438, People's Republic of China \label{164} \and
Astronomy Department, University of Science and Technology of China, Hefei 230026, People's Republic of China \label{165} \and
Department of Physics and Astronomy, Michigan State University, 567 Wilson Rd, East Lansing, MI 48824, USA \label{166} \and
}

\authorrunning{Dahale \& Cho et al.}
\titlerunning{Origin of the ring ellipticity in the black hole images of M87*}

\date{ Received April 21, 2025 / Accepted May 14, 2025 }

\abstract{We investigate the origin of the elliptical ring structure observed in the images of the supermassive black hole \m87, aiming to disentangle contributions from gravitational, astrophysical, and imaging effects. Leveraging the enhanced capabilities of the Event Horizon Telescope (EHT) 2018 array, including improved $(u,v)$-coverage from the Greenland Telescope, we measure the ring's ellipticity using five independent imaging methods, obtaining a consistent average value of $\tau = 0.08_{-0.02}^{+0.03}$ with a position angle $\xi = 50.1_{-7.6}^{+6.2}$ degrees. To interpret this measurement, we compare against General Relativistic Magnetohydrodynamic (GRMHD) simulations spanning a wide range of physical parameters including thermal or non-thermal electron distribution function, spins, and ion-to-electron temperature ratios in both low and high-density regions. We find no statistically significant correlation between spin and ellipticity in GRMHD images. Instead, we identify a correlation between ellipticity and the fraction of non-ring emission, particularly in non-thermal models and models with higher jet emission. These results indicate that the ellipticity measured from the \m87 emission structure is consistent with that expected from simulations of turbulent accretion flows around black holes, where it is dominated by astrophysical effects rather than gravitational ones. Future high-resolution imaging, including space very long baseline interferometry and long-term monitoring, will be essential to isolate gravitational signatures from astrophysical effects.}


\maketitle

\section{Introduction} \label{sec:introduction}

The Event Horizon Telescope (EHT) Collaboration published the first image of a black hole shadow of the supermassive black hole (SMBH) at the center of the giant elliptical galaxy M~87, featuring a distinctive ring-like structure \citep{EHTC2019a,EHTC2019b,EHTC2019c,EHTC2019d,EHTC2019e,EHTC2019f}. In the context of general relativity (GR), the standard usage of the term ``black hole shadow'' is defined as the appearance of a black hole illuminated from all directions, including from behind the observer \citep[e.g., ][]{Falcke2000}, corresponding to the interior of the so-called ``critical curve'' formed by photon trajectories asymptotically approaching bound photon orbits \citep{Gralla_2019}. The ring-like structure in EHT images is primarily a ``direct image'' ($n=0$ emission, where $n$ is the number of half-orbits) which consists of photons from the accretion flow that are strongly lensed by the black hole's gravity but complete zero half-orbits around it before reaching the observer \citep[e.g.,][]{Gralla_2019, Johnson_2020}. The ``photon ring'' is an infinite series of self-similar subrings of light from photons that complete one or more half-orbits ($n\ge1$) around the black hole before reaching the observer \citep{Johnson_2020}. GR predicts that the critical curve is nearly circular for low inclination angles, such as the $\sim17^\circ$ inclination estimated for the M~87 black hole, \m87 \citep{Mertens2016}. Because the black hole spin introduces asymmetry in the shape of the critical curve, if the EHT can provide observational access to the critical curve then measurement of its shape is a pathway to measurement of spin. In this paper, we use ``gravitational ellipticity'' to refer to the shape distortion of the critical curve. For \m87, spin-induced shadow ellipticity is expected to reach up to $\sim$ 0.02 for a spin parameter $a \sim 0.94$ and inclination $i=17^{\circ}$ \citep[e.g., Figure 7, ][]{Johnson_2020}. Moreover, gravitational effects such as displacement of the inner shadow relative to the photon ring, dependent on black hole spin and inclination can also contribute to observed asymmetries \citep[e.g.,][]{Gralla_2019, Chael2021}. Some exotic spacetimes could produce even larger distortions \citep[e.g., Johannsen–Psaltis Metric in Figure 5 of ][]{Younsi2023}.

However, very long baseline interferometry (VLBI) observations do not directly resolve the shadow itself but rather an observed emission structure that appears ring-like due to synchrotron radiation from plasma near the photon orbit \citep[e.g.,][hereafter \citetalias{EHTC2019e}]{EHTC2019e}. This observed shape, which we refer to as the ``emission ellipticity'', can deviate from circularity due to asymmetric plasma distributions. For instance, emission from turbulent flows can introduce ring ellipticity \citep[e.g.,][]{Tiede2022, Tiede2024}.  Finally, we note that limited $(u,v)$-coverage, noise, and algorithmic choices can introduce asymmetries or artifacts in the reconstructed image, even if the underlying source is circular. In this study, we aim to disentangle the contributions to ellipticity by systematically comparing these two sources: gravitational ellipticity and emission ellipticity. Our goal is to assess the degree to which the observed ellipticity in \m87 EHT images is a result of fundamental spacetime properties, astrophysical factors, or artifacts introduced by imaging algorithms.

In the EHT~2017 results of \m87, the observed ring-like structure appeared with $\sim$ zero ellipticity, with a diameter of $42\pm3$\,\textmu as.  While ellipticity was measured in the reconstructed images ($\sim0.05-0.06$), no interpretation or calibration was performed \citep[see Figure 18 in][]{EHTC2019f}. Subsequent analysis by \citet{Tiede2022} demonstrated that images reconstructed using the best set of imaging parameter combinations, so called Top Set, of \ehtim could not reliably recover ellipticity, often favoring circular rings and yielding an upper limit of ellipticity $\sim0.3$. This limitation was primarily due to sparse $(u,v)$-coverage, particularly in the north-south direction, as well as the Top Set imaging parameter combinations that were not fully optimized for elliptical models.  Later, \citet{Tiede2024} reported an \m87 ring ellipticity of $0.09^{+0.07}_{-0.06}$ using \themis, a Bayesian imaging approach, which was consistent with General Relativistic Magnetohydrodynamic (GRMHD) simulations. These results raised important questions about the origins of the ellipticity in the \m87 images and the effectiveness of different imaging methods in accurately recovering it.

\begin{figure}
    \centering
    \includegraphics[width=\columnwidth]{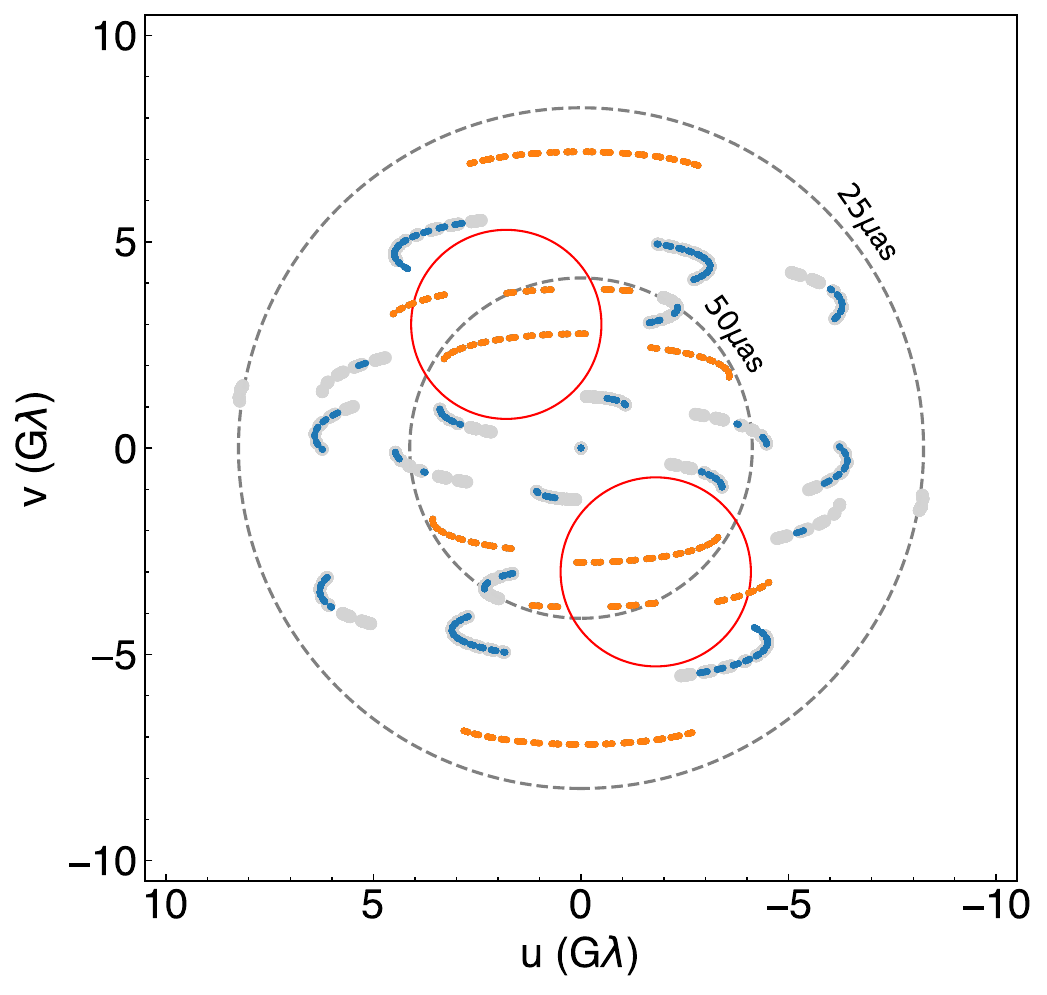}
    \caption{The $(u,v)$-coverage of EHT~2017 (April 10; gray) and EHT~2018 (April 21; blue and orange) observations. Both are at 229.1~GHz, which corresponds to hi-band and band~4 in EHT~2017 and 2018, respectively. 
    Orange points highlight the GLT baselines. Red circles show the coverage gaps in 2017, and dashed circles mark the 25 and 50\,\textmu as resolution.}
    \label{fig:coverage}
\end{figure}

The 2018 observations confirmed the persistent structure of the \m87 black hole shadow with a consistent ring diameter of $43.3^{+1.5}_{-3.1}$\,\textmu as, consistent with the 2017 results \citep[][hereafter \citetalias{EHTC2024a}]{EHTC2024a}. However, annual changes in brightness asymmetry were observed with the position angle shifting from about $180^\circ$ in 2017 to $210^\circ$ in 2018 which may be attributed to turbulence in the accretion flow. Changes in the brightness asymmetry were previously reported by \citet{Wielgus_2020}. The addition of the Greenland Telescope (GLT; \citealt{GLT}) in 2018 substantially improved $(u,v)$-coverage (Fig.~\ref{fig:coverage}), particularly in the north-south direction, leading to improved image fidelity. We note that the EHT~2018 observations included four frequency bands: two at lower frequencies (band~1 and 2 at 213.1\,GHz and 215.1\,GHz) and two at higher frequencies (band~3 and 4 at 227.1\,GHz and 229.1\,GHz). The GLT participated only in the band~3 and 4, so we focus on data from the higher frequency bands in this study. Among the four observing days, we use April~21, which had the highest number of participating stations.

In this study, we follow the formalism based on \citet{Tiede2024} to measure the ellipticity of the ring-like emission structure in \m87 using the 2018 EHT observations. We begin by evaluating the precision of ellipticity measurements using a Fisher information analysis in Sect.~\ref{sec:fisher}. Next, in Sect.~\ref{sec:geometric_test}, we refine the Top Set imaging parameter combinations for the Regularized Maximum Likelihood (RML) and deconvolution imaging methods from \citetalias{EHTC2024a} using elliptical crescent models. Additionally, we test a broader range of elliptical crescent models with various ellipticities and ellipticity position angles to assess any biases in the imaging methods using the data with the new 2018 $(u,v)$-coverage. In Sect.~\ref{sec:m87}, we apply the same imaging pipelines (and parameter combinations) to the EHT~2018 \m87 data to measure the ellipticity of the \m87 emission ring. We then compare these results with those obtained from GRMHD model reconstructions. Finally, in Sect.~\ref{sec:theoretical}, we investigate the origin of the observed ellipticity by comparing our results with theoretical models, followed by a summary and conclusions in Sect.~\ref{sec:conclusions}.

\section{Fisher information analysis} \label{sec:fisher}

\begin{figure*}
    \centering
    \includegraphics[width=\textwidth]{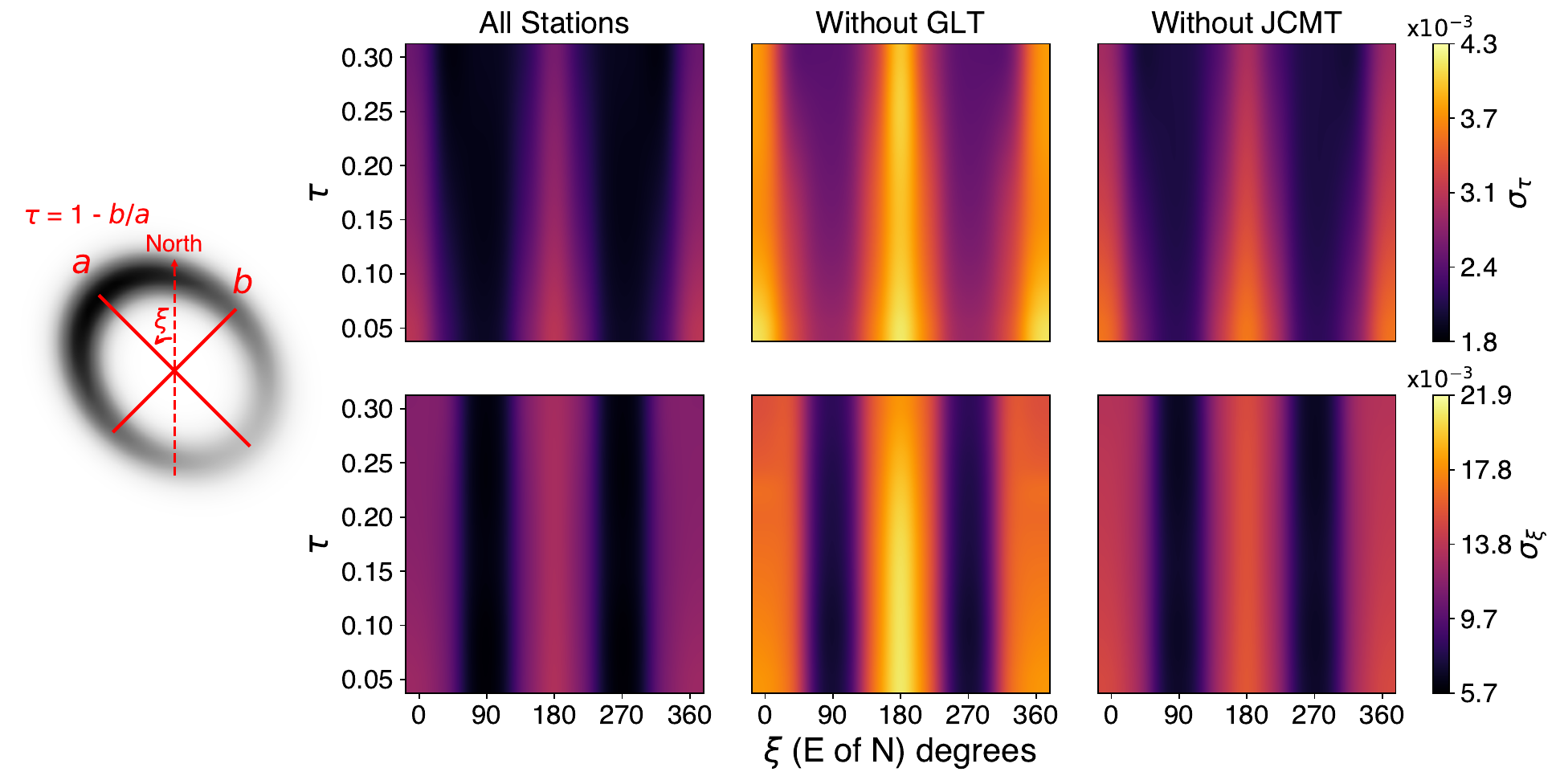}
    \caption{Definition of parameters $\tau$, $\xi$ for a m-ring model, along with their respective marginalized uncertainties using Fisher information analysis. (left) A visualization of a stretched m-ring model with parameters $\tau=0.1$, $\xi=45^{\circ}$, a brightness asymmetry  of 0.23 and a diameter of $d=46$\,\textmu as, blurred with Gaussian kernel with full width at half maximum $10$\,\textmu as. The brightness position angle, $\eta=45^{\circ}$, is aligned with $\xi$.  (right) Maps of marginalized uncertainties in the parameters $\tau$ (top row) and $\xi$ (bottom row) for m-ring models with various values of $\tau$ and $\xi$, derived using Fisher information analysis on \m87 band~4 data.}
    \label{fig:fisher}
\end{figure*}

The Fisher information matrix quantifies the amount of information that an observed data carry about some model parameters $\theta$. The Fisher information matrix is given by,

\begin{equation}
\label{eq:fisher1}
F_{ij} = \mathbb{E} \left[ \frac{\partial \ln \mathcal{L}(\theta)}{\partial \theta_i} \frac{\partial \ln \mathcal{L}(\theta)}{\partial \theta_j} \right],
\end{equation}

where $\mathcal{L}(\theta)$ is the likelihood function of the data conditioned on the parameters $\theta$. The terms $\theta_i$ and $\theta_j$ are elements of the parameter vector $\theta$. The expectation $\mathbb{E}$ is taken with respect to the probability distribution of the data. For independent Gaussian-distributed data with variance $\sigma^2$, the Fisher matrix simplifies to

\begin{equation}
\label{eq:fisher2}
F_{ij} = \sum_k \frac{1}{\sigma_k^2} \left( \frac{\partial V_k}{\partial \theta_i} \frac{\partial V_k^*}{\partial \theta_j} + \frac{\partial V_k^*}{\partial \theta_i} \frac{\partial V_k}{\partial \theta_j} \right),
\end{equation}

where $V_k$ represents the observed complex visibilities. The inverse of the Fisher information matrix gives the covariance matrix of the parameter estimates. Hence, $\Sigma = F^{-1}$, where $\Sigma_{ij}$ represents the covariance between parameters $\theta_i$ and $\theta_j$. We assume a Gaussian posterior distribution for the parameters, where the standard deviations are given by the square root of the diagonal elements of the covariance matrix,

\begin{equation}
\label{eq:fisher4}
\sigma_i = \sqrt{\Sigma_{ii}}\,, 
\end{equation}

which represents the uncertainty in the estimation of each parameter $\theta_i$. 

Given the improved EHT array in 2018, as shown in Fig.~\ref{fig:coverage}, we estimate the precision with which the EHT can measure ellipticity using Fisher information analysis. 

For this analysis, we will use an extension of the m-ring model from \cite{Johnson_2020}. This model has a simple analytic form in both the image and visibility domains, with analytic gradients. This model has also been used for the feature extraction done in \citet[][hereafter \citetalias{EHTC2019d}]{EHTC2019d} and \citetalias{EHTC2024a}, making it useful for physical interpretation. The m-ring model consists of a thin ring with nonuniform brightness in azimuthal directions given by a Fourier series. In polar coordinates $(\rho, \varphi)$, it is defined as,

\begin{equation}
    \mathcal{I}(\rho, \varphi) = \frac{S}{\pi d} \delta\left(\rho - \frac{d}{2}\right) \sum_{k=-m}^{m} \beta_k e^{i k \varphi},
    \label{eq:mring}
\end{equation}

where $S$ is the total flux density of the ring, $d$ is the diameter of the ring, $\delta$ is the Dirac delta function. The coefficients satisfy $\beta_{-k} = \beta_k^*$ for a real image, and we set $\beta_0 = 1$ to ensure that $S > 0$. The parameter $m$ represents azimuthal order of the m-ring. A finite-width m-ring is obtained by convolving Eq.~\ref{eq:mring} with a Gaussian of FWHM $\alpha$. This blurred m-ring is given by

\begin{align}
\mathcal{I}(\rho, \varphi; \alpha) &= \frac{4\ln 2}{\pi\alpha^2}S\exp{\left(-\frac{4\ln 2}{\alpha^2}(\rho^2+d^2/4)\right)} \notag \\
&\quad\times \sum_{k=-m}^{m} \beta_k I_k\left(4\ln 2\frac{\rho d}{2\alpha^2}\right)e^{ik\varphi}\,, 
\end{align}

where $I_k$ denotes the $k$-th modified Bessel function of the first kind \citep{Roelofs_2023}. A stretched m-ring with ellipticity $\tau$, rotated by an ellipticity position angle $\xi$, and width $\alpha$ is given by $\mathcal{I}((1 - \tau\cos(2(\varphi-\xi))\rho,\varphi; \alpha)$.

We use a first-order (i.e., $m=1$) stretched m-ring model as shown in Fig.~\ref{fig:fisher}, as it is the simplest case \citep[e.g.,][]{Tiede2022, Tiede2024}. The ellipticity parameter $\tau$ is defined as $\tau=1-b/a$, where $a$ and $b$ are the major and minor axes of the ellipse, respectively. The ellipticity position angle $\xi$ represents the angle of the major axis $a$, measured counterclockwise from the North (East of North), as shown in Fig.~\ref{fig:fisher}. For this analysis, we keep the brightness position angle (PA), $\eta$, aligned with the ellipticity PA, $\xi$. Hence, we define $\xi$ over the full range of 0 to 360 degrees. We estimate the precision with which the parameters of this model, fitted to data on April 21, 2018 at band~4, can be recovered. We employ the above Fisher information approach implemented within the \texttt{ngEHTforecast} package\footnote{ \href{https://github.com/aeb/ngEHTforecast}{https://github.com/aeb/ngEHTforecast}, accessed with the git commit \texttt{115bf73}.}. This method does not explicitly fit the m-ring model to the data. Instead, it performs a second-order expansion of the logarithmic probability density around the best-fit location, providing an estimate of the uncertainty of each of fitted parameter. The parameter precision estimates assume that the fitting process utilizes complex visibilities as input data, with broad priors imposed on the station gain amplitudes and phases for each scan \citep{Pesce2022}.

For this analysis, we test m-ring models with $\tau$ ranging from $0$ to $0.3$ and $\xi$ ranging from 0 to 360 degrees. The diameter of the thin elliptical m-rings is  $d=\sqrt{ab} = 46$\,\textmu as, which is then blurred by a $10$\,\textmu as circular Gaussian \citep{Tiede2022}. We assume a brightness asymmetry $\beta_1=0.23$ for the m-ring \citepalias[see Table 7 in ][]{EHTC2024a}.
Fig.~\ref{fig:fisher} shows the marginalized uncertainties $\sigma_{\tau}$ and $\sigma_{\xi}$, calculated using Eq.~\ref{eq:fisher4}, for the parameters $\tau$ and $\xi$, respectively. We compute $\sigma_{\tau}$ and $\sigma_{\xi}$ for data with and without GLT to assess its relevance in $(u,v)$-coverage. As seen in the Fig.~\ref{fig:fisher}, $\sigma_{\tau}$ and $\sigma_{\xi}$ are approximately three to five times larger for the data without GLT. This demonstrates that the additional $(u,v)$-coverage provided by GLT baselines enhances the precision in constraining $\tau$ and $\xi$. Furthermore, $\sigma_{\tau}$ and $\sigma_{\xi}$ are nearly three times larger for the models with North-South alignment (i.e., $\xi=0^{\circ}$ or $180^{\circ}$). For models with the same $\xi$ but different $\tau$,  $\sigma_{\tau}$ and $\sigma_{\xi}$ remain approximately constant. During the 2017 and 2018 EHT campaigns the James Clerk Maxwell Telescope (JCMT) only recorded a single polarization feed, which could have contributed to systematic polarization leakage in 2018. To evaluate this issue, we perform a similar analysis for data without JCMT. As shown in Fig.~\ref{fig:fisher}, removing JCMT has a minimal effect ($<10\%$) on the marginalized uncertainties of $\tau$ and $\xi$. Therefore, JCMT data is retained for the remainder of the analysis.

\section{Geometric tests} \label{sec:geometric_test}

Before analyzing the \m87 data, two sets of tests with geometric models were conducted to evaluate the accuracy of the imaging and feature extraction methods in recovering the truth values and assessing potential biases in the imaging methods. We use geometric models with various ellipticity, $\tau$, and the position angle of the ellipse's major axis, $\xi$ (north to east), to check potential biases depending on $(u,v)$-coverage. The first test selects the Top Set imaging parameter combinations of RML and CLEAN methods using four elliptical crescent models ($m=1$) of $\tau=0.187$ and $\xi=[0^\circ, 45^\circ, 90^\circ, 315^\circ]$. The values of $\tau$ and $\xi$ are chosen to be consistent with \citet{Tiede2022}. The second test evaluates the ellipticity feature extraction using 42 elliptical crescent models ($m=1$) with $\tau=[0.0, 0.05, 0.1, 0.15, 0.2, 0.25, 0.3]$ and $\xi=[0^\circ, 60^\circ, 120^\circ, 180^\circ, 240^\circ, 300^\circ]$.  We use \ehtim to generate synthetic data using the geometric models listed above. Before generating the synthetic data, we also add a milli-arcsec scale Gaussian to these geometric models. This Gaussian mimics jet emission on scales of milliarcsecond to arcsecond, to which short intra-site baselines of the EHT are sensitive \citepalias{EHTC2019d}. We add station gain corruptions derived from \m87 and thermal noise to mimic the real observational data \citepalias{EHTC2024a}. Synthetic data with random gain corruptions are also tested using Bayesian imaging, where the posterior distribution of the gain parameters is estimated.

Imaging is performed using both forward and inverse modeling techniques. The forward modeling consists of RML and Bayesian methods, while the inverse modeling employs a CLEAN-based deconvolution method. For RML imaging, we use \ehtim \citep{Chael_2016, Chael_2018, Chael_2019} and \smili \citep{Akiyama_2017a, Akiyama_2017b, Akiyama_2019}. For Bayesian imaging, we use \comrade \citep{Tiede2022_Comrade} and \themis \citep{Broderick_2020a, Broderick_2020b}. For CLEAN-based deconvolution, we use \difmap \citep{Shepherd_1997, Shepherd_2011}. A more detailed explanation of each imaging method is provided in \citetalias{EHTC2019d} and \citetalias{EHTC2024a} (see Appendix~\ref{sec:changes}, for updates to the Bayesian imaging methods). We note that while \citetalias{EHTC2024a} has employed both imaging methods and visibility domain model fitting, we focus on the imaging methods in this study.

\subsection{Imaging parameter selection with the elliptical crescent models}

\begin{figure}
    \centering
    \includegraphics[width=\columnwidth]{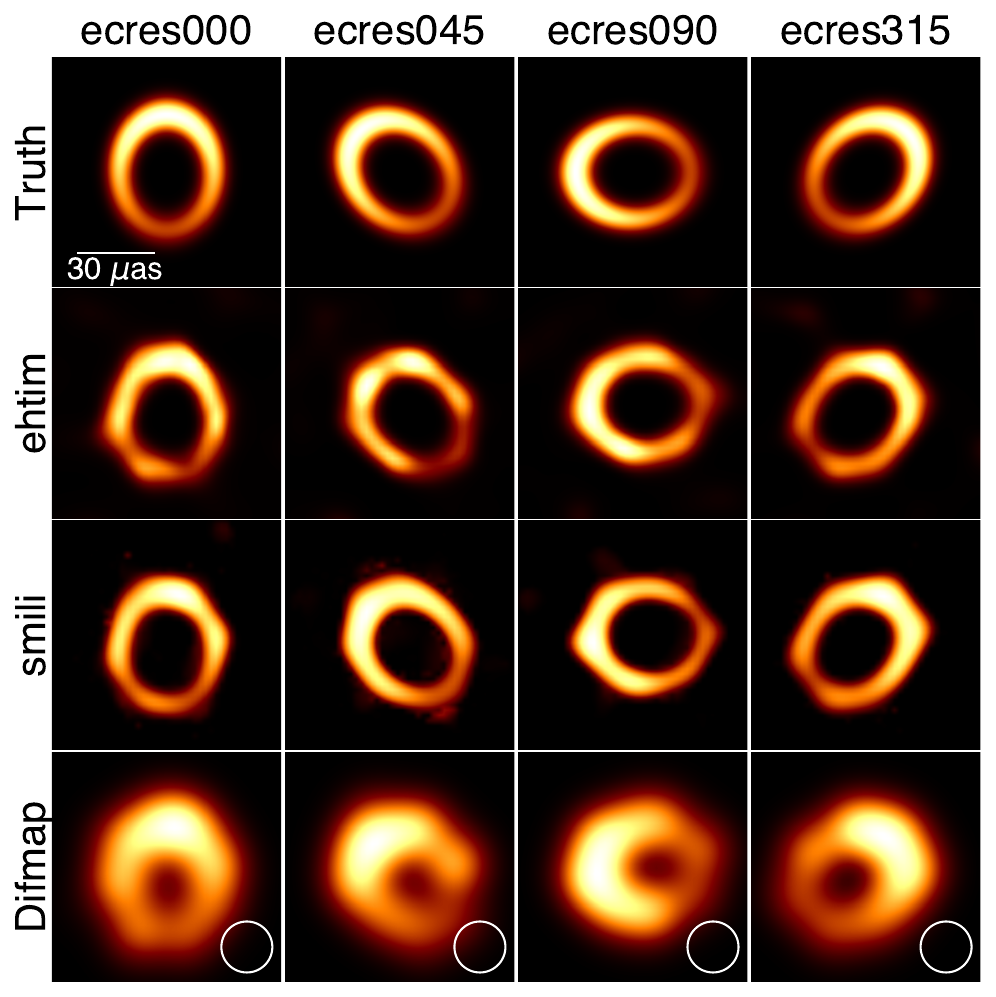}
    \caption{The elliptical crescent models with different ellipse position angle. From left to right, \texttt{ecres000}, \texttt{ecres045}, \texttt{ecres090}, \texttt{ecres315}. From top to bottom, ground truth images, fiducial images from \ehtim, \smili, and \difmap for band~4. \difmap images are presented with a beam convolution of 20\,\textmu as circular Gaussian \citepalias{EHTC2024a}.}
    \label{fig:geometric_ecres_training}
\end{figure}

The data set for the first test is used to sub-select the Top Set imaging parameter combinations for the RML and CLEAN methods. Bayesian methods do not require a parameter survey, and thus, the geometric models used in this step were not tested. Imaging results from the RML and CLEAN methods depend on a set of parameters, which are determined by various imaging assumptions, including hyperparameters and optimization choices.  Each combination of parameters can yield slightly different image morphology and fit quality to the data. Therefore, it is necessary to survey different parameter combinations and select those that provide the best fit to the data and most closely reproduce the ground truth image (if from synthetic data), referred to as the Top Set. For \ehtim, \smili, and \difmap, the Top Set was previously selected in \citetalias{EHTC2024a} based on four geometric models (\texttt{cres180}, \texttt{dblsrc}, \texttt{disk}, \texttt{ring}), using data from \m87. The number of Top Set parameter combinations varies across methods due to differences in their parameter space. Additionally, variations between bands arise from different (systematic) uncertainties inherent in each dataset (see \citetalias{EHTC2024a}). However, it remains untested whether the Top Set parameters are valid for different elliptical structures. To investigate this, we select a new Top Set by imaging four additional geometric models of elliptical crescents:  \texttt{ecres000}, \texttt{ecres045}, \texttt{ecres090}, \texttt{ecres315} (Fig.~\ref{fig:geometric_ecres_training}, top). The imaging survey is performed over the original Top Set from \citetalias{EHTC2024a} meaning the new Top Set corresponds to a subset of the original.

The Top Set selection is based on two metrics: (i) the normalized cross correlation (\nxcorr) between the reconstructed and ground truth images for synthetic data, and (ii) the reduced $\chi^{2}$ on the real \m87 data. Since the latter is already satisfied in the original Top Set (i.e., $\chi^{2}<2$), the new Top Set is selected based solely on the \nxcorr of the elliptical crescent geometric models. The \nxcorr cutoff is determined in the same manner as described in \citetalias{EHTC2019d} and \citetalias{EHTC2024a}, by convolving the ground truth image with the effective resolution from the longest baseline ($\sim 24$\,\textmu as; see Fig.~\ref{fig:coverage}). Then the cutoff value is determined as 0.75 for both band~3 and 4, with no variations across the models.

As a result, several parameter combinations passed the thresholds, demonstrating their ability to reconstruct the elliptical crescent structure while distinguishing the structural position angle. Table~\ref{tab:Top Set} summarizes the number of new Top Set parameter combinations for each pipeline and band compared to the original Top Set. The images reconstructed using the fiducial parameters are shown in Fig.~\ref{fig:geometric_ecres_training}.

\begin{table}
    \centering
    \caption{The number of new Top Set parameter combinations.}
    \resizebox{\columnwidth}{!}{
    \begin{tabular}{cccc}
        \hline\hline
         Band & \ehtim & \smili & \difmap \\
         \hline
         band~3 & 860/874 (98\%) & 4429/5333 (83\%) & 189/303 (62\%) \\
         band~4 & 1332/1469 (91\%) & 3457/5108 (68\%) & 215/465 (46\%) \\
         \hline
    \end{tabular}}
    \tablefoot{This table shows the number of new Top Set parameter combinations compared to the original Top Set. The number ratio relative to the original Top Set is given in parentheses.} 

    \label{tab:Top Set}
\end{table}

\subsection{Evaluation with the elliptical crescent models} \label{sec:validation}

After selecting the new Top Set parameters for the RML (\ehtim and \smili) and CLEAN (\difmap) methods, we perform additional imaging of the geometric models using all methods for performance evaluation. The Bayesian methods, \themis and \comrade, do not require a Top Set selection since their only hyperparameters are field-of-view and the number of pixels. Therefore, we directly perform the geometric tests with these two methods. For these models, $\xi$ coincides with the position angle of the brightest spot (see Appendix~\ref{sec:extratests} for tests with different alignments). This approach aims to identify specific cases where ellipticity or position angle models are not well recovered due to ($u,v$)-coverage limitations. To measure ellipticity and the ellipticity position angle, we use a stretched m-ring template from \texttt{VIDA} \citep{Tiede2022b}. For all feature extraction in this work, we use m-ring of order four in azimuth and order one in width, following \cite{Tiede2022}.

\begin{figure*}
    \centering
    \includegraphics[width=\textwidth]{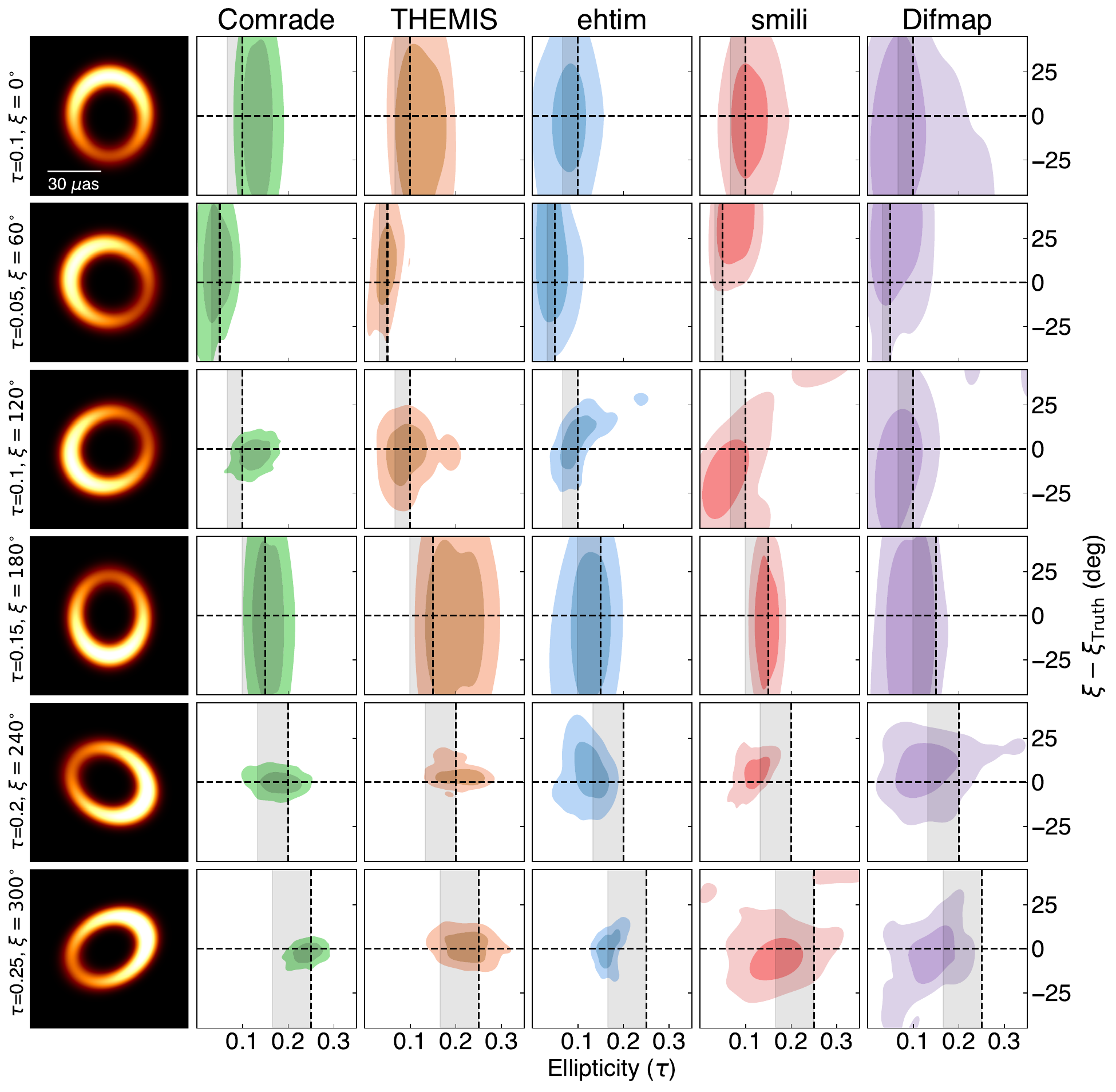}
    \caption{A subset ellipticity measurements for geometric models, 6 out of 42, for all imaging methods. (From left to right) ground truth image of a given geometric model, $\tau-\xi$ distribution using \comrade, \themis, \ehtim, \smili, and \difmap for band 4 synthetic data. Each row corresponds to a different geometric model, with the true model shown in the left-most column. The vertical and horizontal dashed lines in each panel of $\tau-\xi$ distribution indicate the true values. The contours denote $68\%$ and $95\%$ confidence intervals. The vertical dashed line shows $\tau$ measured with VIDA for the ground truth model. The gray shaded region spans the range of $\tau$ values for the ground truth model, from the unconvolved case to convolution of a circular Gaussian of $20$\,\textmu as FWHM.}
    \label{fig:ecres_grid}
\end{figure*}

Fig.~\ref{fig:ecres_grid} presents a subset of the measured $\tau-\xi$ distributions from the respective geometric model reconstructions for different imaging pipelines. The results indicate that $\xi$ is less constrained at $0^\circ$ and $180^\circ$ due to relatively poorer ($u,v$)-coverage in these directions, as expected from the Fisher information analysis (Sect.~\ref{sec:fisher}). The measured ellipticity is still influenced by the underlying angular resolution. For instance, convolving the images and models with $5$\,\textmu as circular Gaussian reduces the ellipticity measurement of $0.1$ by $\sim3$\,\% (see Appendix~\ref{sec:resolution}, for more discussions about the resolution effect on measured ring features).  In our results, the resolution limit of \difmap is given as $\sim20$\,\textmu as \citepalias{EHTC2024a}, while the forward modeling results from RML methods including Bayesian approaches can achieve super-resolution. Therefore, measured ellipticities from reconstructed images that are up to $\sim30\,\%$ lower than the true value are considered acceptable (gray shaded, vertical area in Fig.~\ref{fig:ecres_grid}; see also Fig.~\ref{fig:blur}). As a result, all imaging methods successfully recover the true $\tau$ and $\xi$ values within its own resolution limit. However, it is worth noting that, ignoring the resolution effects, the RML and deconvolution imaging methods tend to underestimate the ellipticity for models with extreme ellipticities in all 42 tests.

\section{Ring ellipticity of M87* and its comparison with GRMHD simulation snapshots} \label{sec:m87}

\begin{table*}
    \centering
    \caption{Measured ellipticity and position angle for EHT~2018 \m87 images.}
    {\renewcommand{\arraystretch}{1.5}
    \begin{tabular}{ccccccccc}
        \hline\hline
        & Band &  \comrade &  \themis &  \ehtim &  \smili & \difmap & Average & Band Average\\
        \hline
        \multirow{2}{*}{$\tau$}& band 3& $0.07^{+0.04}_{-0.03}$ & $0.05^{+0.02}_{-0.03}$ & $0.07^{+0.03}_{-0.03}$ &  $0.09^{+0.04}_{-0.03}$& $0.10^{+0.04}_{-0.04}$ & $0.07^{+0.03}_{-0.03}$ & \multirow{2}{*}{$0.08^{+0.03}_{-0.02}$}\\
        & band 4 & $0.09^{+0.04}_{-0.03}$ & $0.09^{+0.01}_{-0.01}$ & $0.06^{+0.02}_{-0.02}$ &  $0.07^{+0.02}_{-0.02}$ & $0.05^{+0.03}_{-0.02}$ & $0.08^{+0.03}_{-0.02}$ & \\ \hline
        \multirow{2}{*}{$\xi$}& band 3 & $44.1^{+10.5}_{-12.8}$ & $48.3^{+4.0}_{-4.3}$ & $56.0^{+9.8}_{-7.2}$ &  $66.6^{+2.3}_{-8.9}$& $54.2^{+6.2}_{-4.0}$ & $53.5^{+6.0}_{-6.2}$ & \multirow{2}{*}{$50.1^{+6.2}_{-7.6}$}\\
        & band 4 & $40.0^{+11.2}_{-18.3}$ & $38.9^{+6.0}_{-4.1}$ & $48.4^{+10.6}_{-19.7}$ &  $50.4^{+4.9}_{-11.3}$ & $52.4^{+6.5}_{-10.7}$ & $44.0^{+6.7}_{-9.7}$ &\\
        \hline
    \end{tabular}
    }
    \tablefoot{Ellipticity, $\tau$ (in rows labeled $\tau$), and its position angle, $\xi$ in degrees (in rows labeled $\xi$), measured from the EHT~2018 \m87 images. The position angle $\xi$ is measured East of North. For each method and band, the main value reported is the median, and the error range represents the $1\sigma$ uncertainty around the median. The ``Average'' column provides the weighted median and $1\sigma$ error across methods for each band. The ``Band Average'' column gives the overall weighted median and $1\sigma$ error across all methods and both bands for $\tau$ and $\xi$ respectively.}
    \label{tab:m87_ellipticity}
\end{table*}

\begin{figure*}
    \centering
    \includegraphics[width=\textwidth]{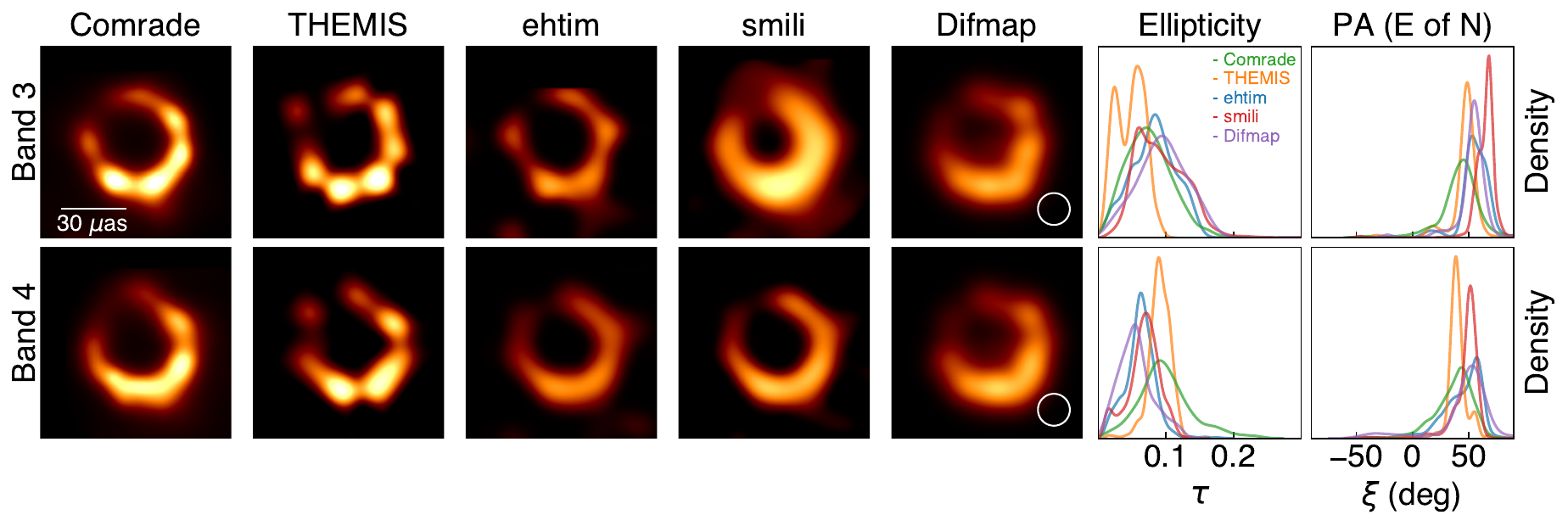}
    \caption{EHT~2018 \m87 images and ellipticity ($\tau$) and ellipticity position angle ($\xi$) distributions. (left to right) Images from each imaging method (\comrade, \themis, \ehtim, \smili, and \difmap). For \comrade and \themis, mean posterior images are shown. Fiducial images from the respective Top Sets are shown for the other methods. The right two panels show the distributions of $\tau$ and $\xi$ for each method. Top and bottom rows show results for band 3 and band 4, respectively. Colors correspond to each method (see legend).} 
    \label{fig:m87_ellipticity}
\end{figure*}

\begin{figure*}
    \centering
    \includegraphics[width=\textwidth]{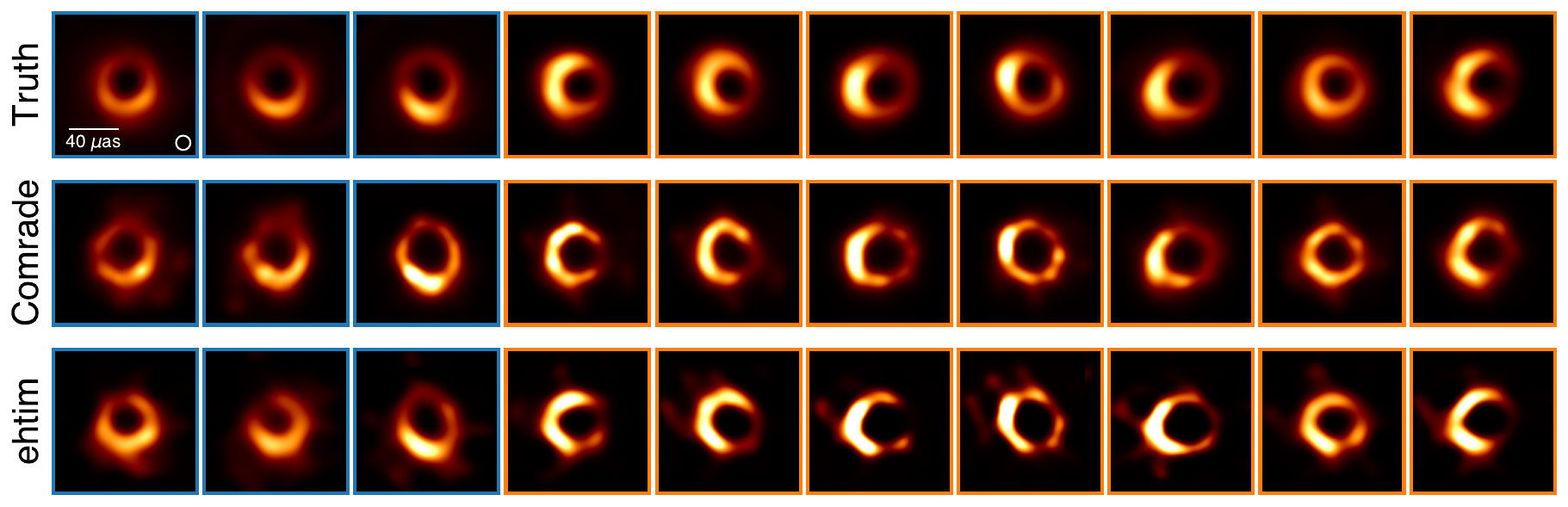}
    \caption{A subset of GRMHD images (10 out of 100 models) from \kharma and \bhac libraries: 
    (from top to bottom) ground truth, reconstructed images by \comrade (mean image) and \ehtim (one random from Top Set) for band~4. Thermal models from \kharma marked in orange and non-thermal models are marked in blue. The ground truths are shown after blurring with a 12\,\textmu as circular Gaussian.}
    \label{fig:grmhd_truth}
\end{figure*}

\subsection{Ring ellipticity of M87*}

Following the geometric tests that validated the imaging pipelines and their parameter combinations for ellipticity measurement, they were applied to the \m87 data\footnote{Data accessed from \url{https://datacommons.cyverse.org/browse/iplant/home/shared/commons_repo/curated/EHTC_M87-2018_Mar2024/}} (see Fig.~\ref{fig:m87_ellipticity}). The imaging results correspond to a subset of images in \citetalias{EHTC2024a} for RML and CLEAN methods (Table~\ref{tab:Top Set}), while remaining consistent for Bayesian methods. Ring features were extracted from the images using \texttt{VIDA}\footnote{A m-ring model of first-order in width and fourth-order in azimuth template was used, consistent with the template used in the geometric tests (see also, \citealt{Tiede2022}).}, as summarized in Table~\ref{tab:m87_ellipticity}. The measured ellipticities are consistent across all imaging approaches and at both frequency bands 3 and 4, yielding an average ellipticity of $\tau=0.08^{+0.03}_{-0.02}$. The average position angle of the ellipse is $\xi=50.1^{+6.2}_{-7.6}$ degrees, consistent across different imaging methods within 2$\,\sigma$. The average across all methods and bands is computed by taking weighted median and $1\sigma$ error. Given the relatively better performance of Bayesian imaging methods, as described in Sect.~\ref{sec:geometric_test} and shown in Fig.~\ref{fig:ecres_grid}, we compute the average $\tau = 0.08 ^{+0.02}_{-0.02}$ and $\xi = 44.4^{+5.8}_{-6.4}$ only from these methods. These averages are in good agreement with the values obtained by averaging over all methods. Notably, the direction of $\xi$ is approximately aligned with the angle of brightest spot on the ring, $\sim200-230^\circ$ \citepalias{EHTC2024a}.

The measured $\tau$ is also consistent with \citet{Tiede2022}, who reported $\tau=0-0.3$ and inferred accretion turbulence as the dominant source of the measured ellipticity, and with \citet{Tiede2024} who reported $\tau=0.09^{+0.07}_{-0.06}$. We note that while this work uses 2018 EHT observations, the above comparisons are made with the works that use 2017 EHT observations. The uncertainties in our results are reduced compared to previous studies from the 2017 EHT data, owing to the improved $(u,v)$-coverage in the 2018~EHT.

\subsection{Comparison with GRMHD simulations} \label{sec:grmhd}

With the results, we apply the same imaging and feature extraction methods to GRMHD models with different physical parameters to investigate the underlying physical dependencies.  For this purpose, we used two GRMHD libraries of \kharma \citep{Prather_2021} and \bhac \citep{Porth_2017}. Thermal electron distribution (Maxwell-Jüttner distribution) models were drawn from \kharma, while non-thermal electron distribution (kappa distribution) models were sourced from \bhac \citep{Fromm2022, Osorio2022}. Out of the 299 GRMHD models that were considered in \citetalias{EHTC2025a}, we selected 218 models with an outflow power exceeding $10^{42}$\,erg/s (\citetalias{EHTC2019a}, \citetalias{EHTC2019e}, \citetalias{EHTC2025a}). After the outflow threshold was applied, we selected 100 models for imaging and feature extraction, of which roughly 75\,\% were thermal models and the remainder non-thermal. There were 18 models for each spin value of $-0.94$, $-0.5$, $+0.5$, and $+0.94$ from the \kharma thermal models, resulting in 72 thermal models in total. The remaining 28 models were chosen from the \bhac non-thermal models that met the threshold, selecting 7 models for each spin. Since black hole spin and inclination determine a displacement of the inner shadow that can manifest as non-circularity \citep[e.g.,][]{Tiede2022}, this sample can investigate the potential correlation between spin and ellipticity. Among the 100 selected GRMHD models, 83 are strongly magnetized magnetically arrested disk (MAD) models and 17 are weakly magnetized standard and normal evolution (SANE) models. Random snapshots were taken from each model and scaled to the best-fit mass based on snapshot scoring implemented in \citetalias{EHTC2025a}. The scaling factor is a ratio of observed to simulated mass-to-distance ratios.

\begin{figure}
    \centering
    \includegraphics[width=\columnwidth]{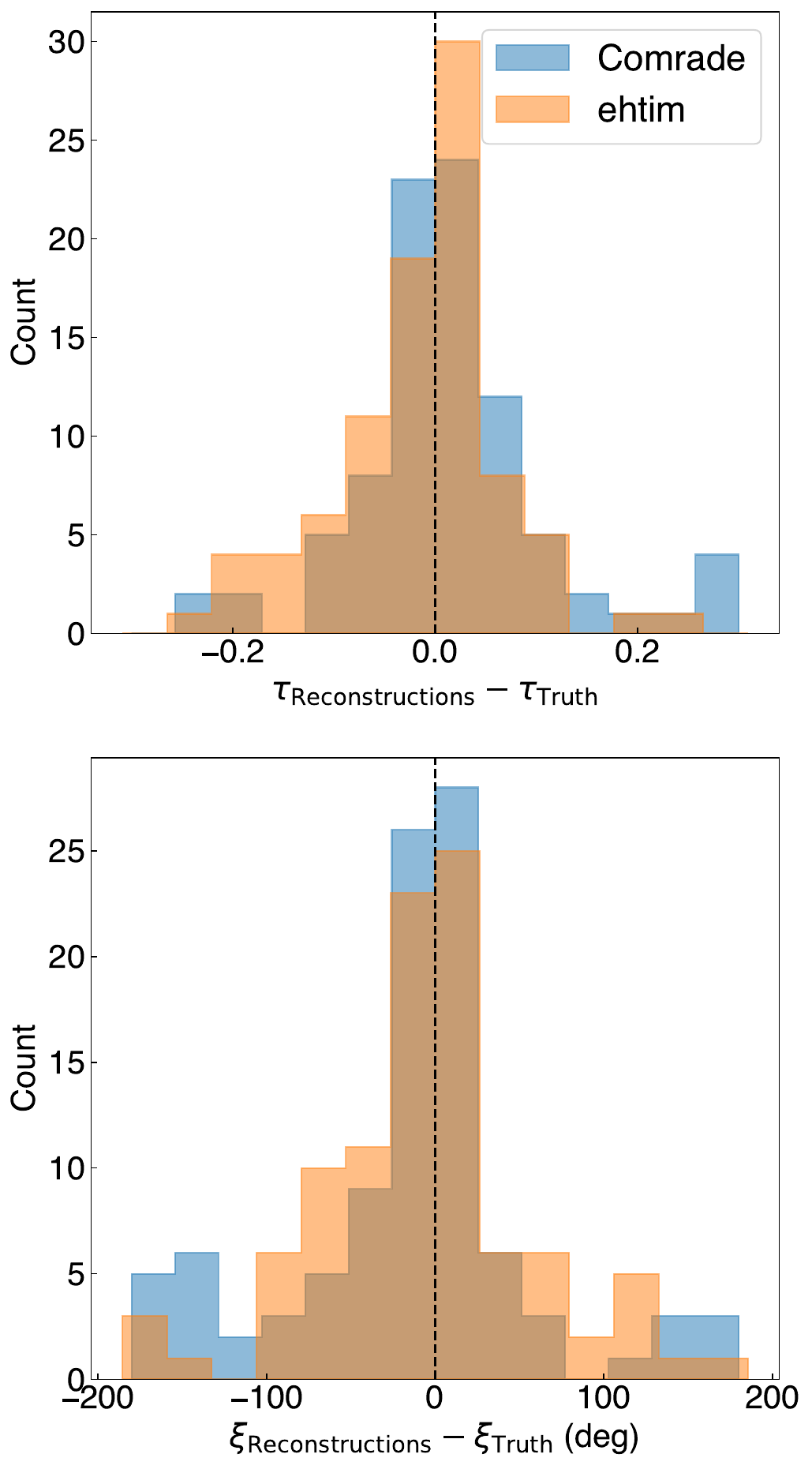}
    \caption{Histograms of the differences in $\tau$ (top) and $\xi$ (bottom) between reconstructed and true GRMHD values. Results from \comrade\ (blue) and \ehtim\ (orange) are shown, both centered around zero for each parameter.} 
    \label{fig:grmhd_residuals}
\end{figure}

The scaled snapshots were then used to generate synthetic data following the same procedure described in Sect.~\ref{sec:geometric_test}. The synthetic data were imaged using \ehtim and \comrade, as representatives of RML and Bayesian methods, respectively. As presented in Fig.~\ref{fig:grmhd_truth}, the images were consistent with the groundtruth. We then extracted the ellipticities from these images using \texttt{VIDA} and compared them with the true ellipticity. The groundtruth GRMHD snapshot images were convolved with 12\,\textmu as circular Gaussian to take resolution effect into account. The 12\,\textmu as size corresponds to the obtained super-resolution from \comrade and \ehtim, that is estimated by comparing \nxcorr between the original and blurred groundtruths images with different sizes of circular Gaussian kernel. This is an average from results for all 100 GRMHD models and is consistent between \comrade and \ehtim. Fig.~\ref{fig:grmhd_residuals} shows the differences of $\tau$ and $\xi$ between reconstructed images (from \ehtim and \comrade) and the true values that are centered at zero in both parameters. This suggests that the observed ellipticity and its angle are real and not an artifact of the imaging process or instrumental limitations. The broader spread in $\xi$ is due to models with low ellipticity, where the orientation angle is naturally more difficult to constrain.

\begin{figure*}
    \centering
    \includegraphics[width=\textwidth]{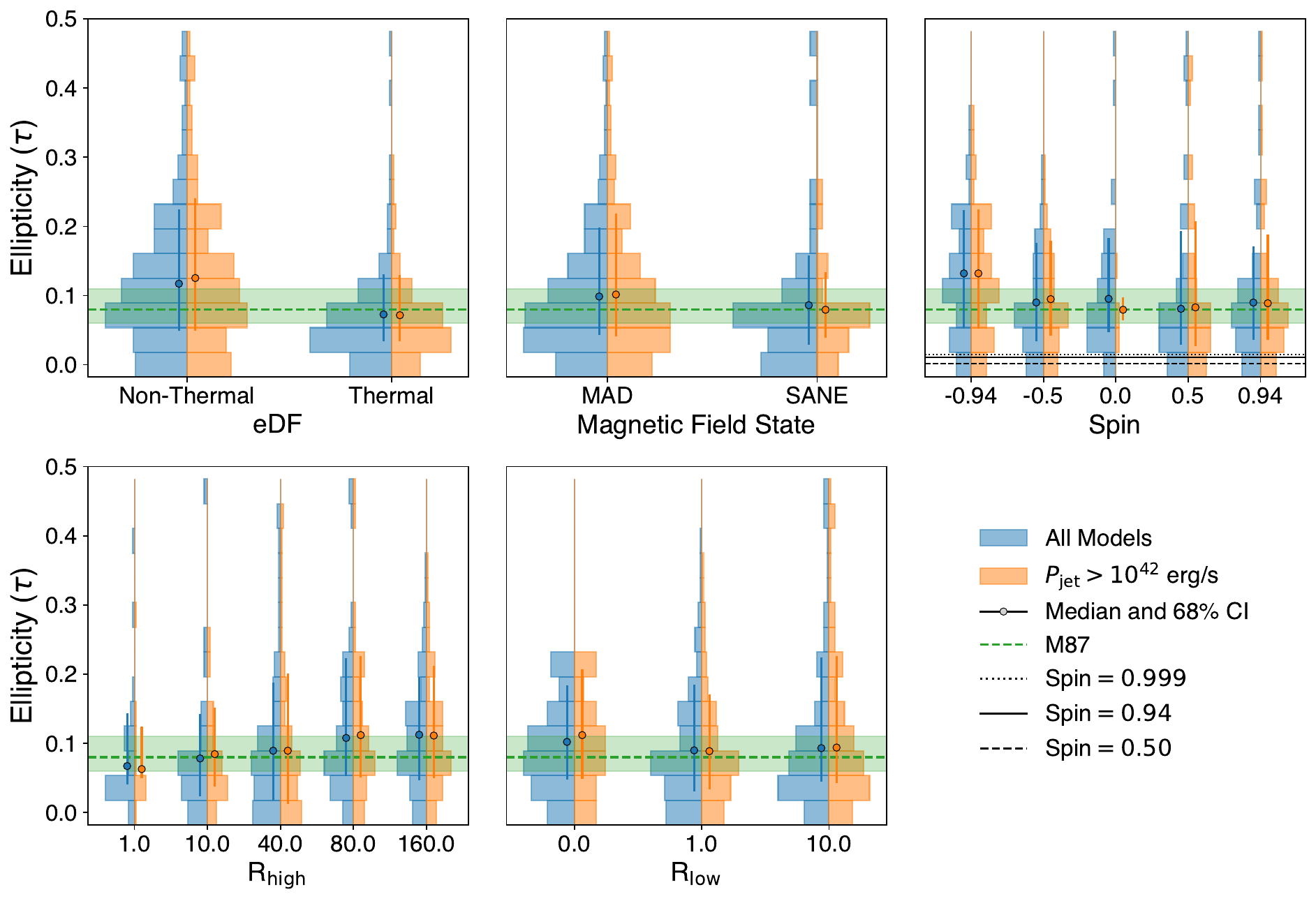}
    \caption{Ellipticity of truth GRMHD models (blurred with 12\,\textmu as Gaussian) for different GRMHD parameters. All 299 models are shown in blue and the models that pass the jet power criteria are shown in orange. (Note that while plotting we have clubbed together the non-thermal models with the same parameters but different non-thermal emission fraction ($\epsilon$).) The over-plotted on histograms is the median value and the 68\% confidence interval. The ellipticity of M87 is shown in green dashed line with its error range shown in green shaded region. The top-rightmost panel, the ellipticity of the critical curve for a Kerr black hole \citep[see Figure 7,][]{Johnson_2020} for different spins (inclination=$17^{\circ}$) are plotted to compare with the histograms of the respective spins. The spin=$0.5$ case is shown by black dashed line, spin=$+0.94$ case by black solid line and spin=$+0.999$ case by black dotted line.}
    \label{fig:grmhd_all}
\end{figure*}

\begin{figure*}
    \centering
    \includegraphics[width=\textwidth]{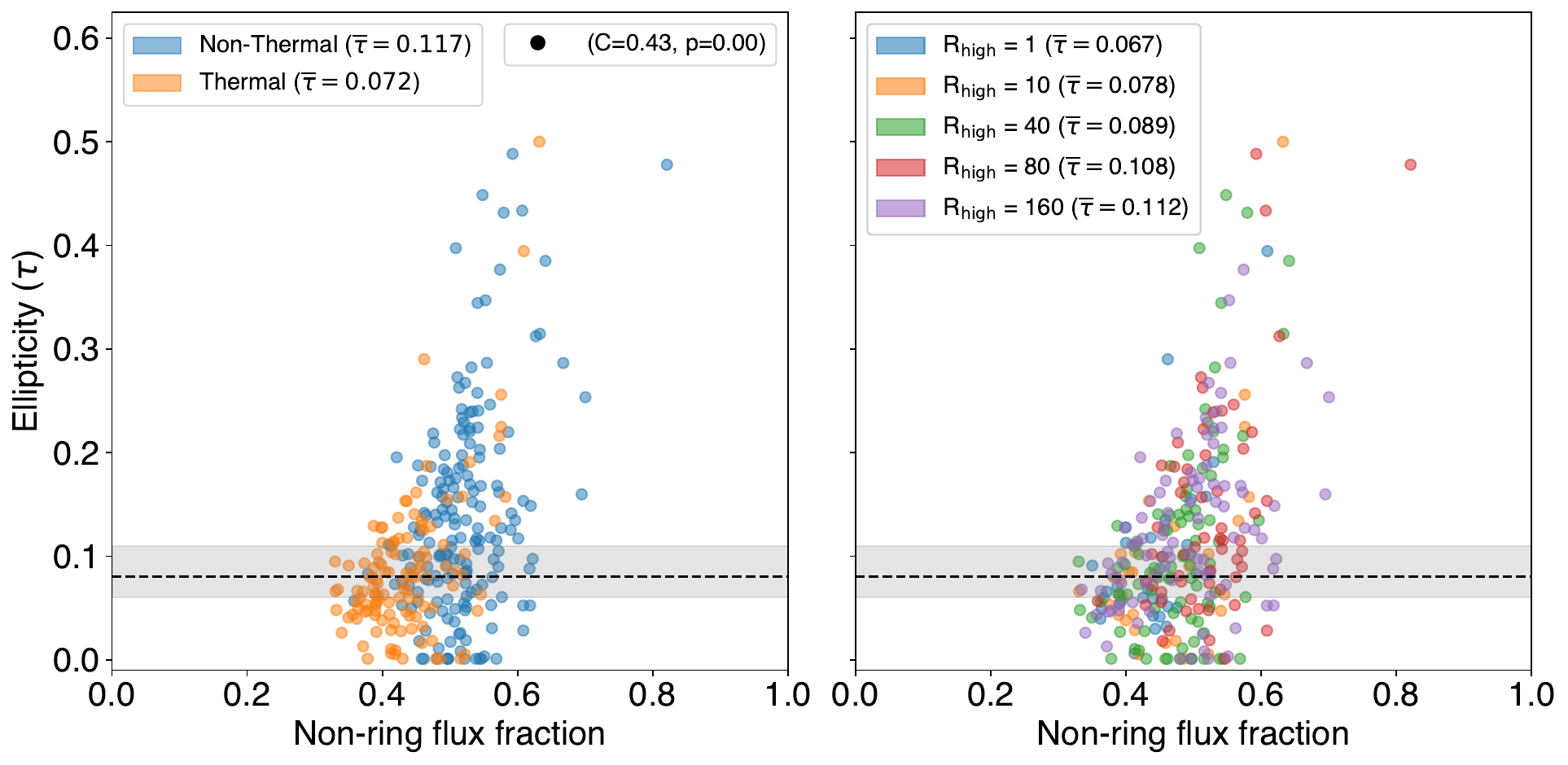}
    \caption{The ellipticity of all 299 GRMHD truths (convolved with 12\,\textmu as Gaussian) compared with the non-ring flux fraction. The left and right panels are the same but the colors are different for (non-)thermal models (left) and \rhigh values (right). The ellipticity of M87 is shown in black dashed line with its error range shown in gray shaded region. The median $\tau$ for each case is shown in the legend.}
    \label{fig:correlations}
\end{figure*}

\begin{table*}
    \centering
     \caption{Spearman correlation for ellipticity of GRMHD truth models and their physical parameters.}
    \resizebox{\textwidth}{!}{
    \begin{tabular}{cccccc}
        \hline \hline
        Models                                           & eDF  & Magnetic Field State & Spin      & \rhigh    & \rlow     \\ \hline
        All                                   &  \ccg $C= -0.31, p=0.05$            &  $C=-0.09, p=0.44$ &  $C=-0.12, p=0.22$ &   \ccg $C=0.20, p=0.04$ &  $C=-0.02, p=0.50$ \\
        $P_{\text{jet}} > P_{0}$      &  \ccg $C= -0.35, p=0.03$            &  $C=-0.17, p=0.26$ &  $C=-0.13, p=0.18$ &   \ccg $C=0.22, p=0.03$ &  $C=-0.02, p=0.48$  \\ \hline
    \end{tabular}
    }
    \tablefoot{Spearman correlation coefficient $C$ and its associated $p$-value for the ellipticity of GRMHD truth models and their physical parameters. These were computed using a bootstrapping method \citep[e.g.,][]{Curran_2014, Cheng_2023}. A $p$-value $\le 0.05$ indicates a rejection of the null hypothesis that the parameters are not correlated. Cases considered correlated ($p \le 0.05$) are highlighted in green. $P_{0} =10^{42}$ erg/s is the jet power threshold applied to the models.}
    \label{tab:grmhd_correlation}
\end{table*}

\begin{figure*}
    \centering
    \includegraphics[width=\textwidth]{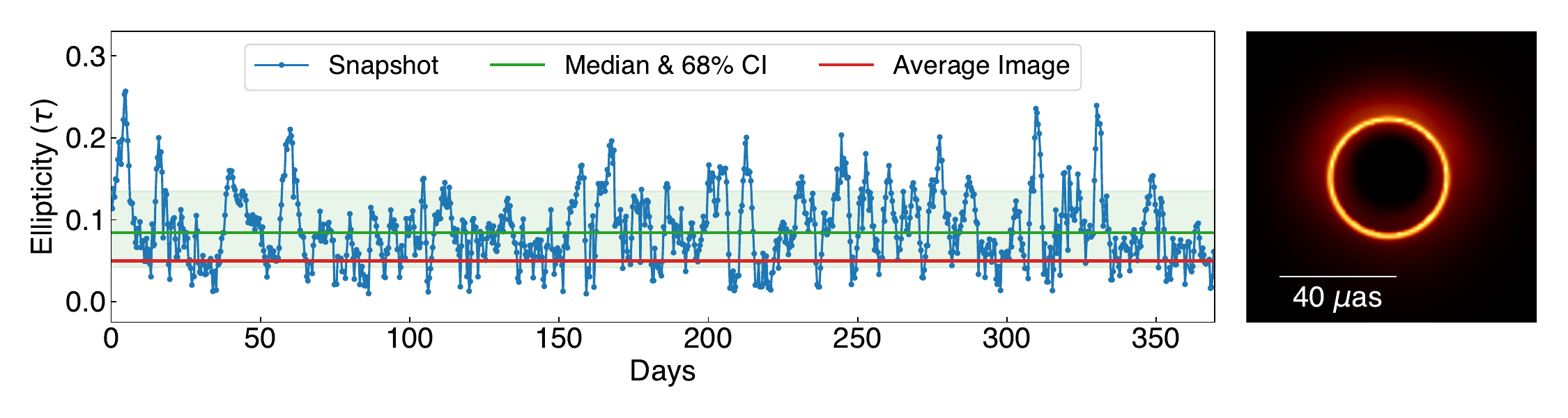}
    \caption{Ellipticities of 1000 snapshots spanning $\sim 370$ days for a GRMHD model with the following parameters: magnetic field configuration = MAD, thermal eDF, black hole spin = $-0.5$, \rhigh $= 160$, \rlow $= 1$, and inclination angle = $17^{\circ}$. The ellipticity of each snapshot is measured after blurring with a $12$\,\textmu as Gaussian (blue). The median and $68\%$ confidence interval of the time series are shown with a solid green line and shaded region, respectively. The ellipticity of the time-averaged image (right) is shown with a red line.}
    \label{fig:time_series}
\end{figure*}

\section{Origin of the M87* ring ellipticity} \label{sec:theoretical}

As introduced in Sec.~\ref{sec:introduction}, observed ellipticity in EHT images can arise from two main sources: (i) gravitational ellipticity due to spacetime curvature, and (ii) emission ellipticity from astrophysical structure. In this section, we focus on disentangling the contributions using GRMHD simulations with varying physical parameters. We utilize all 299 GRMHD models described in Sect.~\ref{sec:grmhd}, which span a wide range of parameters, including black hole spin, electron distribution function (eDF) (thermal and non-thermal), magnetic field states (SANE and MAD), and ion-to-electron temperature ratios in both low- and high-density regions (\rlow and \rhigh, respectively; see, \citealt{Moscibrodzka2016} for definitions). For each model, we measure the image ellipticity from the corresponding groundtruth GRMHD snapshot, scaled to the best-fit mass and blurred with a $12$\,\textmu as Gaussian as described in Sect.~\ref{sec:grmhd}. Fig.~\ref{fig:grmhd_all} shows that the observed \m87 ellipticity is consistent with the distribution of all 299 GRMHD models, falling within their median and 68\% confidence intervals.

GRMHD simulations are uniquely suited to probe both gravitational ellipticity and emission ellipticity. The gravitational ellipticity arises from the curvature of spacetime near the event horizon and is directly influenced by black hole spin. For instance, a Kerr black hole with spin $a = 0.94-0.999$ and viewed at an inclination of $17^\circ$ can produce a gravitational ellipticity of $\sim 0.02$ (see Fig.~\ref{fig:grmhd_all}, top right). This level of distortion cannot explain the measured ellipticity of \m87. Even if we consider a $3^{\circ}$ uncertainty \citep{Mertens2016} on the inclination of \m87, with $20^{\circ}$ inclination and spin $a=0.999$, the maximum ellipticity reaches only $\sim 0.03$ \citep{Johnson_2020}. Moreover, as shown in Table~\ref{tab:grmhd_correlation} and Fig.~\ref{fig:grmhd_all}, there is no statistically significant correlation between spin and the measured ellipticity in full GRMHD images. This indicates that while spin contributes to the gravitational ellipticity, it does not dominate the total observed ellipticity in the images. Table~\ref{tab:grmhd_correlation} also shows no significant correlation between ellipticity and magnetic field state (SANE and MAD), implying that the global magnetic field structure does not have a strong influence on ellipticity. In contrast, Fig.~\ref{fig:grmhd_all} and Table~\ref{tab:grmhd_correlation} suggest that non-thermal models and simulations with higher values of \rhigh are correlated with larger emission ellipticity. Non-thermal models tend to generate more extended or diffuse emission, and increasing \rhigh shifts emission from the disk to the jet region, both effects that increase ellipticity of the ring in the image.

To quantify the role of asymmetric non-ring emission, we define a ``non-ring flux fraction'' by subtracting the best-fit circular Gaussian ring (using \texttt{VIDA}) from each GRMHD image and setting negative residuals to zero. The ratio of the remaining positive flux to the total flux defines the non-ring flux fraction. We compute Spearman correlation coefficients between this quantity and the measured image ellipticity, using a bootstrapping method to account for sample variance \citep[e.g.,][]{Curran_2014, Cheng_2023}. Fig.~\ref{fig:correlations} reveals a positive correlation between the non-ring flux fraction and ellipticity. We note that upon visual inspection, it does not appear that the outliers appear from transient flux eruption events. This supports the conclusion that emission ellipticity, driven by turbulent accretion structures outside the ring, is the dominant contributor to ellipticity in GRMHD images. These results suggest that the observed ellipticity in the \m87 image is naturally explained by emission ellipticity arising from astrophysical effects such as turbulent accretion flow.

\section{Summary and conclusions} \label{sec:conclusions}

In this study, we measured the ellipticity of the emission ring in the black hole images of \m87 using the EHT~2018 observations. Fisher information analysis was employed to first assess the feasibility of ellipticity measurements, after the addition of the GLT, which filled gaps in the north-south $(u,v)$-coverage. This analysis shows that the inclusion of GLT improves the precision in constraining ellipticity parameters by $\sim3-5$ times. With the improved $(u,v)$-coverage compared to the EHT~2017, we then managed to extract the ellipticity successfully from five different methods that cross-compare the results. This is the first method-wide confirmation of the ellipticity measurement.

For imaging with the RML and deconvolution methods, the imaging parameter combinations were sub-selected from the original ones in \citetalias{EHTC2024a}, based on imaging results for four elliptical geometric models. This step is not required for Bayesian imaging methods such as \comrade and \themis. All imaging methods (and parameters) were then evaluated using various geometric elliptical ring models with differing position angles and ellipticities, and they successfully recovered the ground truth values in most cases.

Applying these approaches to the M87 data, we measured the ellipticity to be $\tau = 0.08^{+0.03}_{-0.02}$ on average, with consistency across all imaging methods and previous findings from EHT~2017 \citep{Tiede2024}. In addition, the position angle of the ellipse is measured as $\xi=50.1^{+6.2}_{-7.6}$ degrees, indicating the ring structure is slightly elongated along an axis that is roughly aligned with the brightest spot on the ring, $\sim200-230^\circ$ \citepalias{EHTC2024a}. We also note that while \cite{Kim2025} reported $\tau = 0.06 \pm 0.04$ for the image of M87 at 86\,GHz, which is consistent with our measurements at 230\,GHz, this is an astrophysical effect at 86\,GHz. Comparison with GRMHD simulations first confirms that the measured ellipticity is real, but no strong constraints on the parameters including the spin are yet found. However, the parameters providing more non-ring flux (that is, non-thermal emission and higher \rhigh) tend to reproduce larger ellipticity. This is confirmed by comparison with the non-ring flux fraction to the measured ellipticity. In line with this, to explain the measured ellipticity of M87, the astrophysical effects such as turbulent accretion flow are required. It is worth noting that the additional non-circularity by the inclined black hole (e.g., tilted accretion disk) or exotic spacetime are not completely ruled out.

With the current ground-based array, our ability to precisely measure the ellipticity of the emission ring is constrained by the dominance of turbulent astrophysical effects associated with the direct image \citep{Gralla2020a, Johnson_2020}. In the case of \m87, detecting the gravitational influence on ellipticity and hence measuring the spin, requires overcoming these limitations through one of two approaches. The first approach involves continued long-term monitoring of \m87, allowing for temporal averaging of the turbulent effects, thereby enabling the underlying gravitational signature to emerge more clearly. For instance, in Fig.~\ref{fig:time_series}, we present the measured ellipticities of 1000 snapshots spanning $\sim 370$ days for one of the GRMHD models, along with the ellipticity of the time-averaged image in the right panel. The average image clearly reveals the photon ring, while the turbulent accretion flow is averaged out. This example illustrates that future observations could detect ellipticity arising from gravitational effects, as highlighted by the red line in the left panel.

The second approach entails space VLBI observations, which would provide the necessary angular resolution to detect the photon ring \citep{Gralla_2019, Johnson_2020}. Unlike the direct image, the photon ring is expected to be less affected by astrophysical turbulence, making it a more direct probe of the underlying spacetime structure and gravitational effects near the black hole. We also note that the direct image is more elliptical than the photon ring \citep[see Figure 6 in][]{Gralla2020a}. To better distinguish the relative effects on the ellipticity from different potential origins, higher precision measurements of ellipticities from better angular resolution are necessary. Looking ahead, future observations with higher angular resolution, such as those from the next-generation EHT (ngEHT; \citealt{Johnson_2023, Doeleman_2023}), the Event Horizon Imager (EHI; \citealt{Roelofs2019}), the Terahertz Exploration and Zooming-in for Astrophysics (THEZA; \citealt{Gurvits2022}) and the Black Hole Explorer (BHEX; \citealt{Johnson_2024, Akiyama_2024}), will offer further constraints by resolving finer photon ring structures. As demonstrated by the addition of the GLT to the EHT array, the inclusion of future sites in the EHT and ngEHT arrays will improve the precision of measurements of the ring ellipticity of \m87.

\bibliographystyle{aa}
\bibliography{references}

\appendix

\section{Bayesian imaging methods} \label{sec:changes}

\begin{figure}[ht!]
    \centering
    \includegraphics[width=\columnwidth]{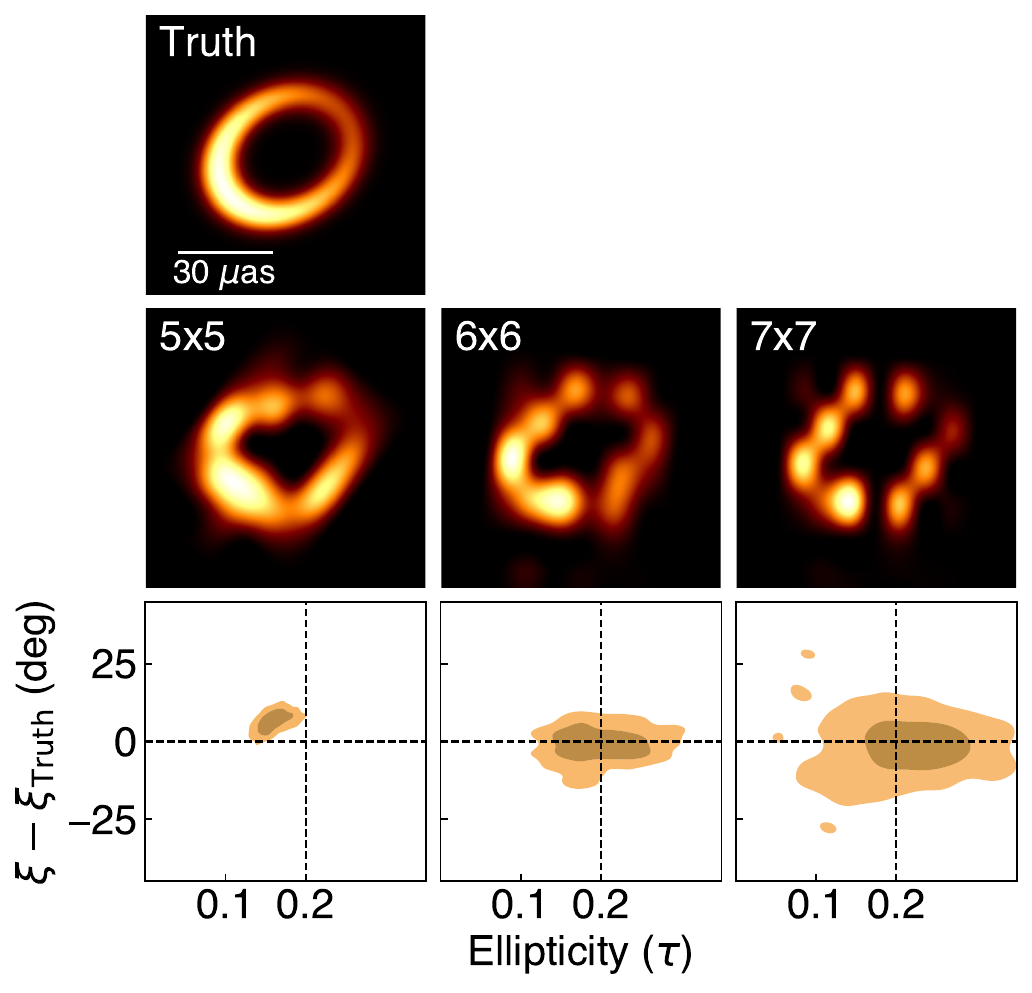}
    \caption{Evaluation of different rasters of \themis through synthetic data tests. The true m-ring model with $\tau$=0.1 and $\xi =120^{\circ}$ is shown on the top. The posterior mean images from \themis for $5\times5$, $6\times6$, and $7\times7$ rasters are shown in the middle. $\tau$ and  $\xi$ - $\xi_{\mathrm{Truth}}$ posteriors are shown at the bottom. The contours are shown for $68\%$ and $95\%$ confidence intervals. The dashed line marks the truth values.}
    \label{fig:raster_test}
\end{figure}

In \citetalias{EHTC2024a}, \themis used a $5\times5$ pixel grid (raster) as it was used in \citetalias{EHTC2019d}. Since this size, may not be suitable for the 2018 EHT coverage or elliptical rings, similar to \cite{Tiede2024}, we performed a small survey of different imaging models, changing the number of pixels and computing the Bayesian evidence to find the optimal raster size. The Bayesian evidence ($Z$) of a model $M$ with parameters $\bm{\theta}$ is given by,

\begin{equation}
    Z(M) = \int \mathcal{L}(\bm{\mathcal{V}} | \bm{\theta}, M)p(\bm{\theta}, M) d\bm{\theta}.
\end{equation}

where $\mathcal{L}$ is the log-likelihood, and $\mathcal{V}$ is the observed data. In a Bayesian setting, the optimal model corresponds to the one with the highest evidence. The Bayesian evidence was computed using thermodynamic integration \citep{lartillot2006computing}, utilizing the parallel tempering scheme from \cite{syed_non-reversible_2019}. Given the computational complexity of running a survey for each set of data, we instead focused on one of the geometric elliptical ring synthetic data sets in Sect.~\ref{sec:geometric_test}. Specifically, we considered the elliptical ring with $\tau=0.1$ and $\xi=120^{\circ}$. The evidence for this survey is given in Tab~\ref{tab:grid_evidences}. According to these values, $6\times6$ raster has the highest evidence. Furthermore, Fig.~\ref{fig:raster_test}, which shows how the true value of $\tau=0.1$ and $\xi=120^{\circ}$ is recovered correctly by the $6\times6$ raster. Hence, throughout the analysis in this paper, we used the $6\times6$ raster for \themis.

\begin{table}[ht!]
    \centering
    \begin{tabular}{lccc}
    \hline \hline
      Raster size &  $5\times5$ & $6\times6$ & $7\times7$\\
      Evidence &  $-2890$ & $-2885$ & $-2889$\\
    \hline
    \end{tabular}
    \caption{Evidence of different raster sizes}
    \label{tab:grid_evidences}
\end{table}

In \citetalias{EHTC2024a}, \comrade imaging used closure phases and visibility amplitudes as the data products, and fit only for gain amplitudes. The prior used for the raster assumed that the pixels are independently distributed (Dirichlet prior). There have been several developments in \comrade since then. In this paper, we used complex visibilities as the data products and fit for both amplitude and phase gains (with reference to one station). The image model is a $64\times64$ raster with a field of view of $200$\,\textmu as. An extended Gaussian component was used to model the emission on milliarcsecond scales. The location, position angle, and fractional flux are the model parameters, while the size is fixed as $1000$\,\textmu as. The total flux of the image plus the Gaussian component was fixed to 1.1\,Jy as assumed in network calibration for the M87* data. For the raster, we use a first-order Gaussian Markov random field (GMRF) prior on the log-ratio transformed pixel intensities. Hence, the pixels are spatially correlated. The GMRF is added to a mean image which is a $40$\,\textmu as Gaussian (size of the Gaussian does not drastically change the reconstructed images). The variance and correlation length of the random field are the hyperparameters. The variance and the correlation length are included as parameters in the model. The amplitude gain priors are the same as used in \citetalias{EHTC2024a}. We allow gains to vary every scan. For the gain phases, the gain phase for ALMA is set to be zero. In the case when ALMA is not present in the scan, we select the next reference station alphabetically. For rest of the gain phase priors, we use von Mises prior with zero mean and a concentration parameter $\pi^{-2}$ (which is essentially a uniform prior on the interval $[-\pi, \pi]$). Sampling is performed using \cite{NUTS}, in the Julia sampling package \texttt{AdvancedHMC.jl}\footnote{\href{https://github.com/TuringLang/AdvancedHMC.jl}{https://github.com/TuringLang/AdvancedHMC.jl}}, in the same way as done in \citetalias{EHTC2024a}.

\section{Additional geometric tests} \label{sec:extratests}

\begin{figure}[ht!]
    \centering
    \includegraphics[width=\columnwidth]{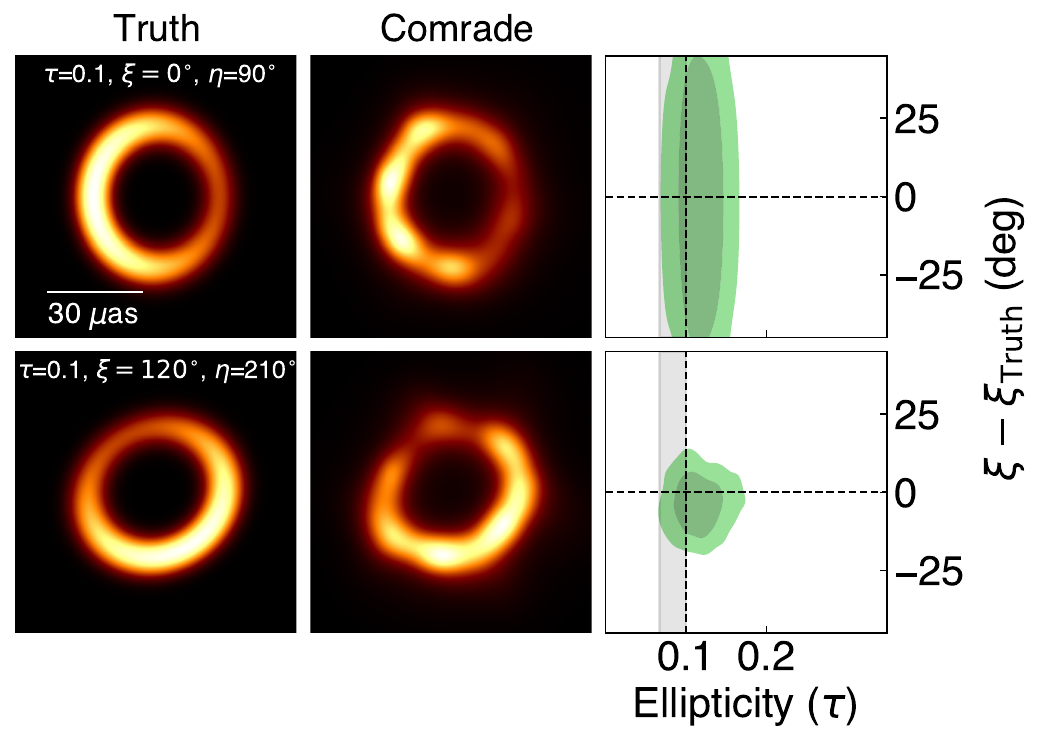}
    \caption{Geometric tests for the cases when brightness PA  $\eta$ is perpendicular to $\xi$. The leftmost column shows the ground truth for the cases $\tau=0.1$, $\xi=0^{\circ}$, $\eta=90^{\circ}$ (top) and $\xi=120^{\circ}$ (bottom). Mean images from \comrade posteriors are shown in the middle. The rightmost column shows measured $\tau$ and $\xi$ posteriors ($68\%$ and $95\%$ confidence contours) compared with true values in dashed lines. The gray shaded region marks the region between $\tau$ of the ground truth model without any convolution and with convolution of a Gaussian up to $20$\,\textmu as size. } 
    \label{fig:brightness_test}
\end{figure}

In Sect.~\ref{sec:geometric_test}, we performed the tests with geometric models where the ellipticity position angle $\xi$ was aligned with the brightness position angle $\eta$. In order to be certain that an offset between position angles, will not introduce any additional biases in measuring ellipticity, we performed this additional test shown in Fig.~\ref{fig:brightness_test}. We chose the $\eta$ to be perpendicular to $\xi$ to test the extreme case. This test was performed with \comrade for all $\tau$ and $\xi$ but Fig.~\ref{fig:brightness_test} shows only the two cases when $\xi=0^{\circ}$ and $\xi=120^{\circ}$. In all cases, \comrade recovers the true $\tau$ correctly. For $\xi$, we see the same pattern as seen in Fig.~\ref{fig:ecres_grid} and as mentioned in Sec.~\ref{sec:geometric_test}. Even in these tests, we found that true $\xi$ can be recovered with narrow posteriors for all cases, except when $\xi$ is aligned North-South, for which we get broad posteriors.

\section{The effect of resolution on the measured ellipticity} \label{sec:resolution}

\texttt{VIDA} stretched m-ring template models the ellipticity of a m-ring by stretching a symmetric m-ring template in $x$-direction and compressing it in $y$-direction. Assuming this definition of measured ellipticity, we want to solve for an analytical equation that relates $\tau$ with a Gaussian blurring kernel. To do so, consider a zero-order symmetric m-ring with flux $F_{0}$ and radius $r_{0}$, $\frac{F_{0}}{2\pi r_{0}}\delta(r-r_{0})$, which has a Fourier transform given by, $F_{0}J_{0}(2\pi r_{0} |u|)$. $J_{0}$ is a zeroth-order Bessel function of the first kind and $u, v$ are the spatial coordinates in the Fourier domain. Stretching the m-ring in the $u$-direction by $\beta$ and compressing in the $v$-direction by $\beta$, we get:

\begin{equation}
    \label{eq:blur3}
    \tilde{\mathcal{I}}(u,v;r_0, \tau) = F_{0}J_{0}\left(2\pi r_{0} \sqrt{\frac{u^2}{1-\tau} +(1-\tau)v^2} \right)
\end{equation}

where $\tilde{\mathcal{I}}(u,v)$ is the intensity profile in the Fourier domain (or the amplitude visibilities at $u,v)$, $\tau=1-b/a = 1- (2r_{0}\beta)/(2r_{0}/\beta) = 1-\beta^2$. If we blur this stretched m-ring with a circular Gaussian of full-width at half maximum (FWHM) $\alpha$, we get,

\begin{equation}
    \label{eq:blur4}
      \tilde{\mathcal{I}}(u,v;r_0, \tau, \alpha) =  \tilde{I}(u,v;r_0, \tau) \times  \exp{\left(\frac{-\pi^2|u|^2\alpha^2}{4\log2}\right)}
\end{equation}

From Eq.~\ref{eq:blur4}, it is not possible to get a $\tau-\alpha$ relation without the $u$ and $v$ dependence. Instead, we used geometric elliptical m-ring models with $\xi=120^{\circ}$ and $\tau_{0}=[0.05, 0.1, 0.15, 0.2, 0.25, 0.3]$ and measured the ellipticity with \texttt{VIDA} by blurring the models with different circular Gaussian FWHM ($\alpha$). We then fitted a Gaussian to each measured $\tau-\alpha$ data (see Eq.~\ref{eq:blur_effect_gaussian}) for all different $\tau_0-\xi$ models.

\begin{equation}
    \label{eq:blur_effect_gaussian}
    \tau = A * e^{-(\alpha-\mu)^2/(2\sigma^2)}
\end{equation}

\begin{figure}[ht!]
    \centering
    \includegraphics[width=\columnwidth]{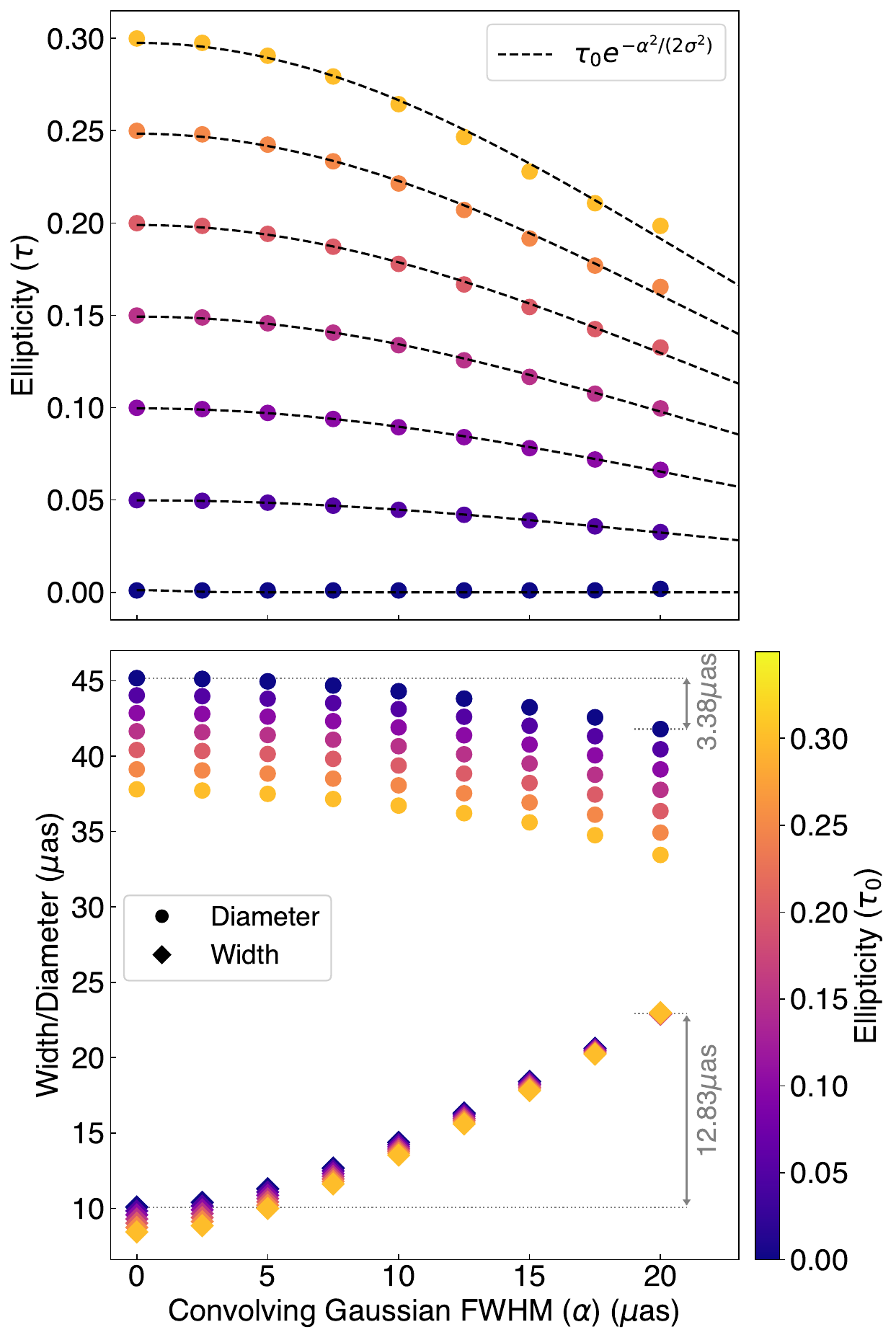}
    \caption{The effect of blurring with different Gaussian FWHM ($\alpha$) on the ring parameters measured by \texttt{VIDA}. (Top) Ellipticity of a stretched m-ring with true $\xi =120^{\circ}$ and $\tau_{0} = [0.05, 0.1, 0.15, 0.2, 0.25, 0.3]$. The black dashed line marks the fitted curves. (Bottom) Diameter (circle) and width (diamond) of a stretched m-ring with the same true $\xi$ and $\tau_{0}$ as on the top.}
    \label{fig:blur}
\end{figure}

When Eq.~\ref{eq:blur_effect_gaussian} is fit to different $\tau-\alpha$ data for the $\tau_{0}$ cases, we get $\mu=0$ and $A=\tau_{0}$ for all the cases as shown in Eq.~\ref{eq:blur_effect}.

\begin{equation}
    \label{eq:blur_effect}
    \tau = \tau_{0} e^{-\alpha^2/(2\sigma^2)}
\end{equation}

where $\sigma$ was measured as $21.5\pm0.5$ \textmu as for all the cases. We note that we have data only up to one $\sigma$ of the fitted Gaussians. The drop in $\tau$ is high when $\tau_{0}$ is high, as shown in Fig.~\ref{fig:blur} (top). The maximum change is seen in the $\tau_{0}=0.3$ case, when it is blurred by $\alpha=20$\,\textmu as, $\tau$ goes from 0.3 to $\sim$ 0.225. When $0.05 < \tau_{0}< 1.0$, given the nominal resolution of $\sim 20$ \textmu as, the change in true ellipticity $\Delta \tau$ is $0.017 \lesssim \Delta\tau \lesssim 0.034$. While the diameter of the m-ring drops by a maximum of $\sim4$\,\textmu as, the width increases by a maximum of $\sim13$\,\textmu as, after blurring with $\alpha=20$\,\textmu as as shown in Fig.~\ref{fig:blur} (bottom).

\section{Acknowledgements}\label{sec:acknowledgements}

\begin{acknowledgements} 
R.D. acknowledges that the project that gave rise to these results received the support of a fellowship from “la Caixa” Foundation (ID 100010434), with the fellowship code LCF/BQ/DI22/11940030. I.C. is supported by the KASI-Yonsei Postdoctoral Fellowship.

The Event Horizon Telescope Collaboration thanks the following organizations and programs: the Academia Sinica; the Academy of Finland (projects 274477, 284495, 312496, 315721); the Agencia Nacional de Investigaci\'{o}n y Desarrollo (ANID), Chile via NCN$19\_058$ (TITANs), Fondecyt 1221421 and BASAL FB210003; the Alexander
von Humboldt Stiftung; an Alfred P. Sloan Research Fellowship; Allegro, the European ALMA Regional Centre node in the Netherlands, the NL astronomy research network NOVA and the astronomy institutes of the University of Amsterdam, Leiden University, and Radboud University; the ALMA North America Development Fund; the Astrophysics and High Energy Physics programme by MCIN (with funding from European Union NextGenerationEU, PRTR-C17I1); the Black Hole Initiative, which is funded by grants from the John Templeton Foundation (60477, 61497, 62286) and the Gordon and Betty Moore Foundation (Grant GBMF-8273) - although the opinions expressed in this work are those of the author and do not necessarily reflect the views of these Foundations; 
the Brinson Foundation; the Canada Research Chairs (CRC) program; Chandra DD7-18089X and TM6-17006X; the China Scholarship Council; the China Postdoctoral Science Foundation fellowships (2020M671266, 2022M712084); ANID through Fondecyt Postdoctorado (project 3250762); Conicyt through Fondecyt Postdoctorado (project 3220195); Consejo Nacional de Humanidades, Ciencia y Tecnología (CONAHCYT, Mexico, projects U0004-246083, U0004-259839, F0003-272050, M0037-279006, F0003-281692, 104497, 275201, 263356, CBF2023-2024-1102, 257435); the Colfuturo Scholarship; the Consejo Superior de Investigaciones Cient\'{i}ficas (grant 2019AEP112); the Delaney Family via the Delaney Family John A. Wheeler Chair at Perimeter Institute; Dirección General de Asuntos del Personal Académico-Universidad Nacional Autónoma de México (DGAPA-UNAM, projects IN112820 and IN108324); the Dutch Research Council (NWO) for the VICI award (grant 639.043.513), the grant OCENW.KLEIN.113, and the Dutch Black Hole Consortium (with project No. NWA 1292.19.202) of the research programme the National Science Agenda; the Dutch National Supercomputers, Cartesius and Snellius  (NWO grant 2021.013); the EACOA Fellowship awarded by the East Asia Core Observatories Association, which consists of the Academia Sinica Institute of Astronomy and Astrophysics, the National Astronomical Observatory of Japan, Center for Astronomical Mega-Science, Chinese Academy of Sciences, and the Korea Astronomy and Space Science Institute; 
the European Research Council (ERC) Synergy Grant ``BlackHoleCam: Imaging the Event Horizon of Black Holes'' (grant 610058) and Synergy Grant ``BlackHolistic:  Colour Movies of Black Holes: Understanding Black Hole Astrophysics from the Event Horizon to Galactic Scales'' (grant 10107164); the European Union Horizon 2020
research and innovation programme under grant agreements RadioNet (No. 730562), M2FINDERS (No. 101018682) and FunFiCO (No. 777740); the European Research Council for advanced grant ``JETSET: Launching, propagation and emission of relativistic jets from binary mergers and across mass scales'' (grant No. 884631); the European Horizon Europe staff exchange (SE) programme HORIZON-MSCA-2021-SE-01 grant NewFunFiCO (No. 10108625); the Horizon ERC Grants 2021 programme under grant agreement No. 101040021; the FAPESP (Funda\c{c}\~ao de Amparo \'a Pesquisa do Estado de S\~ao Paulo) under grant 2021/01183-8; the Fondes de Recherche Nature et Technologies (FRQNT); the Fondo CAS-ANID folio CAS220010; the Generalitat Valenciana (grants APOSTD/2018/177 and  ASFAE/2022/018) and GenT Program (project CIDEGENT/2018/021); the Gordon and Betty Moore Foundation (GBMF-3561, GBMF-5278, GBMF-10423); the Institute for Advanced Study; the ICSC – Centro Nazionale di Ricerca in High Performance Computing, Big Data and Quantum Computing, funded by European Union – NextGenerationEU; the Istituto Nazionale di Fisica Nucleare (INFN) sezione di Napoli, iniziative specifiche
TEONGRAV;  the International Max Planck Research School for Astronomy and Astrophysics at the Universities of Bonn and Cologne; the Italian Ministry of University and Research (MUR)– Project CUP F53D23001260001, funded by the European Union – NextGenerationEU;  DFG research grant ``Jet physics on horizon scales and beyond'' (grant No. 443220636); Joint Columbia/Flatiron Postdoctoral Fellowship (research at the Flatiron Institute is supported by the Simons Foundation); 
the Japan Ministry of Education, Culture, Sports, Science and Technology (MEXT; grant JPMXP1020200109); 
the Japan Society for the Promotion of Science (JSPS) Grant-in-Aid for JSPS Research Fellowship (JP17J08829); the Joint Institute for Computational Fundamental Science, Japan; the Key Research Program of Frontier Sciences, Chinese Academy of Sciences (CAS, grants QYZDJ-SSW-SLH057, QYZDJSSW-SYS008, ZDBS-LY-SLH011); 
the Leverhulme Trust Early Career Research Fellowship; the Max-Planck-Gesellschaft (MPG); the Max Planck Partner Group of the MPG and the
CAS; the MEXT/JSPS KAKENHI (grants 18KK0090, JP21H01137, JP18H03721, JP18K13594, 18K03709, JP19K14761, 18H01245, 25120007, 19H01943, 21H01137, 21H04488, 22H00157, 23K03453); the MICINN Research Projects PID2019-108995GB-C22, PID2022-140888NB-C22; the MIT International Science and Technology Initiatives (MISTI) Funds; 
the Ministry of Science and Technology (MOST) of Taiwan (103-2119-M-001-010-MY2, 105-2112-M-001-025-MY3, 105-2119-M-001-042, 106-2112-M-001-011, 106-2119-M-001-013, 106-2119-M-001-027, 106-2923-M-001-005, 107-2119-M-001-017, 107-2119-M-001-020, 107-2119-M-001-041, 107-2119-M-110-005, 107-2923-M-001-009, 108-2112-M-001-048, 108-2112-M-001-051, 108-2923-M-001-002, 109-2112-M-001-025, 109-2124-M-001-005, 109-2923-M-001-001,
110-2112-M-001-033, 110-2124-M-001-007 and 110-2923-M-001-001); the National Science and Technology Council (NSTC) of Taiwan (111-2124-M-001-005, 112-2124-M-001-014 and  112-2112-M-003-010-MY3); the Ministry of Education (MoE) of Taiwan Yushan Young Scholar Program; the Physics Division, National Center for Theoretical Sciences of Taiwan; the National Aeronautics and Space Administration (NASA, Fermi Guest Investigator grant
80NSSC23K1508, NASA Astrophysics Theory Program grant 80NSSC20K0527, NASA NuSTAR award  80NSSC20K0645); NASA Hubble Fellowship Program Einstein Fellowship; NASA Hubble Fellowship grants HST-HF2-51431.001-A, HST-HF2-51482.001-A, HST-HF2-51539.001-A, HST-HF2-51552.001A awarded by the Space Telescope Science Institute, which is operated by the Association of Universities for Research in Astronomy, Inc., for NASA, under contract NAS5-26555; the National Institute of Natural Sciences (NINS) of Japan; the National Key Research and Development Program of China (grant 2016YFA0400704, 2017YFA0402703, 2016YFA0400702); the National Science and Technology Council (NSTC, grants NSTC 111-2112-M-001 -041, NSTC 111-2124-M-001-005, NSTC 112-2124-M-001-014); the US National Science Foundation (NSF, grants AST-0096454, AST-0352953, AST-0521233, AST-0705062, AST-0905844, AST-0922984, AST-1126433, OIA-1126433, AST-1140030, DGE-1144085, AST-1207704, AST-1207730, AST-1207752, MRI-1228509, OPP-1248097, AST-1310896, AST-1440254,  AST-1555365, AST-1614868, AST-1615796, AST-1715061, AST-1716327,  AST-1726637, 
OISE-1743747, AST-1743747, AST-1816420, AST-1935980, AST-1952099, AST-2034306,  AST-2205908, AST-2307887);  NSF Astronomy and Astrophysics Postdoctoral Fellowship (AST-1903847); the Natural Science Foundation of China (grants 11650110427, 10625314, 11721303, 11725312, 11873028, 11933007, 11991052, 11991053, 12192220, 12192223, 12273022, 12325302, 12303021); the Natural Sciences and Engineering Research Council of Canada (NSERC);  
the National Research Foundation of Korea (the Global PhD Fellowship Grant: grants NRF-2015H1A2A1033752; the Korea Research Fellowship Program: NRF-2015H1D3A1066561; Brain Pool Program: RS-2024-00407499;  Basic Research Support Grant 2019R1F1A1059721, 2021R1A6A3A01086420, 2022R1C1C1005255, 2022R1F1A1075115); Netherlands Research School for Astronomy (NOVA) Virtual Institute of Accretion (VIA) postdoctoral fellowships; NOIRLab, which is managed by the Association of Universities for Research in Astronomy (AURA) under a cooperative agreement with the National Science Foundation; Onsala Space Observatory (OSO) national infrastructure, for the provisioning
of its facilities/observational support (OSO receives funding through the Swedish Research Council under grant 2017-00648);  the Perimeter Institute for Theoretical Physics (research at Perimeter Institute is supported by the Government of Canada through the Department of Innovation, Science and Economic Development and by the Province of Ontario through the Ministry of Research, Innovation and Science); the Portuguese Foundation for Science and Technology (FCT) grants (Individual CEEC program - 5th edition, \url{https://doi.org/10.54499/UIDB/04106/2020}, \url{https://doi.org/10.54499/UIDP/04106/2020}, PTDC/FIS-AST/3041/2020, CERN/FIS-PAR/0024/2021, 2022.04560.PTDC); the Princeton Gravity Initiative; the Spanish Ministerio de Ciencia, Innovaci\'{o}n  y Universidades (grants PID2022-140888NB-C21, PID2022-140888NB-C22, PID2023-147883NB-C21, RYC2023-042988-I); the Severo Ochoa grant CEX2021-001131-S funded by MICIU/AEI/10.13039/501100011033; The European Union’s Horizon Europe research and innovation program under grant agreement No. 101093934 (RADIOBLOCKS); The European Union “NextGenerationEU”, the Recovery, Transformation and Resilience Plan, the CUII of the Andalusian Regional Government and the Spanish CSIC through grant AST22\_00001\_Subproject\_10; ``la Caixa'' Foundation (ID 100010434) through fellowship codes LCF/BQ/DI22/11940027 and LCF/BQ/DI22/11940030; the University of Pretoria for financial aid in the provision of the new Cluster Server nodes and SuperMicro (USA) for a SEEDING GRANT approved toward these nodes in 2020; the Shanghai Municipality orientation program of basic research for international scientists (grant no. 22JC1410600); the Shanghai Pilot Program for Basic Research, Chinese Academy of Science, Shanghai Branch (JCYJ-SHFY-2021-013); the Simons Foundation (grant 00001470); the Spanish Ministry for Science and Innovation grant CEX2021-001131-S funded by MCIN/AEI/10.13039/501100011033; the Spinoza Prize SPI 78-409; the South African Research Chairs Initiative, through the South African Radio Astronomy Observatory (SARAO, grant ID 77948),  which is a facility of the National 
Research Foundation (NRF), an agency of the Department of Science and Innovation (DSI) of South Africa; the Swedish Research Council (VR); the Taplin Fellowship; the Toray Science Foundation; the UK Science and Technology Facilities Council (grant no. ST/X508329/1); the US Department of Energy (USDOE) through the Los Alamos National
Laboratory (operated by Triad National Security, LLC, for the National Nuclear Security Administration of the USDOE, contract 89233218CNA000001); and the YCAA Prize Postdoctoral Fellowship. This work was also supported by the National Research Foundation of Korea (NRF) grant funded by the Korea government (MSIT) (RS-2024-00449206). We acknowledge support from the Coordenação de Aperfeiçoamento de Pessoal de Nível Superior (CAPES) of Brazil through PROEX grant number 88887.845378/2023-00. We acknowledge financial support from Millenium Nucleus NCN23\_002 (TITANs) and Comité Mixto ESO-Chile.

We thank the staff at the participating observatories, correlation centers, and institutions for their enthusiastic support. This paper makes use of the following ALMA data: ADS/JAO.ALMA\#2017.1.00841.V and ADS/JAO.ALMA\#2016.1.01154.V. ALMA is a partnership of the European Southern Observatory (ESO; Europe, representing its member states), NSF, and National Institutes of Natural Sciences of Japan, together with National Research Council (Canada), Ministry of Science and Technology (MOST; Taiwan),
Academia Sinica Institute of Astronomy and Astrophysics (ASIAA; Taiwan), and Korea Astronomy and Space Science Institute (KASI; Republic of Korea), in cooperation with the Republic of Chile. The Joint ALMA Observatory is operated by ESO, Associated Universities, Inc. (AUI)/NRAO, and the National Astronomical Observatory of Japan (NAOJ). The NRAO is a facility of the NSF operated under cooperative agreement by AUI. This research used resources of the Oak Ridge Leadership Computing Facility at the Oak Ridge National Laboratory, which is supported by the Office of Science of the U.S. Department of Energy under contract No. DE-AC05-00OR22725; the ASTROVIVES FEDER infrastructure, with project code IDIFEDER-2021-086; the computing cluster of Shanghai VLBI correlator supported by the Special Fund  for Astronomy from the Ministry of Finance in China; We also thank the Center for Computational Astrophysics, National Astronomical Observatory of Japan. This work was supported by FAPESP (Fundacao de Amparo a Pesquisa do Estado de Sao Paulo) under grant 2021/01183-8.

APEX is a collaboration between the Max-Planck-Institut f{\"u}r Radioastronomie (Germany), ESO, and the Onsala Space Observatory (Sweden). The SMA is a joint project between the SAO and ASIAA and is funded by the Smithsonian Institution and the Academia Sinica. The JCMT is operated by the East Asian Observatory on behalf of the NAOJ, ASIAA, and KASI, as well as the Ministry of Finance of China, Chinese Academy of Sciences, and the National Key Research and Development Program (No. 2017YFA0402700) of China and Natural Science Foundation of China grant 11873028. Additional funding support for the JCMT is provided by the Science and Technologies Facility Council (UK) and participating universities in the UK and Canada. The LMT is a project operated by the Instituto Nacional de Astr\'{o}fisica, \'{O}ptica, y Electr\'{o}nica (Mexico) and the University of Massachusetts at Amherst (USA). The IRAM 30-m telescope on Pico Veleta, Spain is operated by IRAM and supported by CNRS (Centre National de la Recherche Scientifique, France), MPG (Max-Planck-Gesellschaft, Germany), and IGN (Instituto Geogr\'{a}fico Nacional, Spain). The SMT is operated by the Arizona Radio Observatory, a part of the Steward Observatory of the University of Arizona, with financial support of operations from the State of Arizona and financial support for instrumentation development from the NSF. Support for SPT participation in the EHT is provided by the National Science Foundation through award OPP-1852617 to the University of Chicago. Partial support is also provided by the Kavli Institute of Cosmological Physics at the University of Chicago. The SPT hydrogen maser was provided on loan from the GLT, courtesy of ASIAA.

This work used the Extreme Science and Engineering Discovery Environment (XSEDE), supported by NSF grant ACI-1548562, and CyVerse, supported by NSF grants DBI-0735191,
DBI-1265383, and DBI-1743442. XSEDE Stampede2 resource at TACC was allocated through TG-AST170024 and TG-AST080026N. XSEDE JetStream resource at PTI and TACC was allocated through AST170028. This research is part of the Frontera computing project at the Texas Advanced Computing Center through the Frontera Large-Scale Community Partnerships allocation AST20023. Frontera is made possible by National Science Foundation award OAC-1818253. This research was done using services provided by the OSG Consortium~\citep{osg07,osg09}, which is supported by the National Science Foundation award Nos. 2030508 and 1836650.
Additional work used ABACUS2.0, which is part of the eScience center at Southern Denmark University, and the Kultrun Astronomy Hybrid Cluster (projects Conicyt Programa de Astronomia Fondo Quimal QUIMAL170001, Conicyt PIA ACT172033, Fondecyt Iniciacion 11170268, Quimal 220002). Simulations were also performed on the SuperMUC cluster at the LRZ in Garching, on the LOEWE cluster in CSC in Frankfurt, on the HazelHen cluster at the HLRS in Stuttgart, and on the Pi2.0 and Siyuan Mark-I at Shanghai Jiao Tong University. The computer resources of the Finnish IT Center for Science (CSC) and the Finnish Computing  Competence Infrastructure (FCCI) project are acknowledged. This research was enabled in part by support provided by Compute Ontario (\url{http://computeontario.ca}), Calcul Quebec (\url{http://www.calculquebec.ca}), and the Digital Research Alliance of Canada (\url{https://alliancecan.ca/en}).

The EHTC has received generous donations of FPGA chips from Xilinx Inc., under the Xilinx University Program. The EHTC has benefited from technology shared under open-source license by the Collaboration for Astronomy Signal Processing and Electronics Research (CASPER). The EHT project is grateful to T4Science and Microsemi for their
assistance with hydrogen masers. This research has made use of NASA's Astrophysics Data System. We gratefully acknowledge the support provided by the extended
staff of the ALMA, from the inception of the ALMA Phasing Project through the observational campaigns of 2017 and 2018. We would like to thank A. Deller and W. Brisken for EHT-specific support with the use of DiFX. We thank Martin Shepherd for the addition of extra features in the Difmap software  that were used for the CLEAN imaging results presented in this paper. We acknowledge the significance that Maunakea, where the SMA and JCMT EHT stations are located, has for the indigenous Hawaiian people.

\end{acknowledgements}

\end{document}